\documentclass[aps, prb, superscriptaddress, twocolumn]{revtex4-1}
\usepackage{hyperref}
\usepackage{graphicx}
\usepackage{amsmath}
\usepackage{amssymb}
\usepackage{amsfonts}   
\usepackage{multirow}
\usepackage[b]{esvect} 
\usepackage{braket}
\usepackage{color}

\renewcommand{\Im}{\operatorname{Im}}

\newcommand{\bea}{\begin{eqnarray}}
\newcommand{\eea}{\end{eqnarray}}
\newcommand{\beq}{\begin{equation}}
\newcommand{\eeq}{\end{equation}}

\begin{document}
\title{Transiently enhanced interlayer tunneling in optically driven high-$T_c$ superconductors}

\author{Jun-ichi Okamoto}
\email{ojunichi@physnet.uni-hamburg.de}
\affiliation{Zentrum f\"ur Optische Quantentechnologien and Institut f\"ur Laserphysik, Universit\"at Hamburg, 22761 Hamburg, Germany}
\affiliation{The Hamburg Centre for Ultrafast Imaging, Luruper Chaussee 149, 22761 Hamburg, Germany}

\author{Wanzheng Hu}
\affiliation{Max Planck Institute for the Structure and Dynamics of Matter, 22761 Hamburg, Germany}

\author{Andrea Cavalleri}
\affiliation{Max Planck Institute for the Structure and Dynamics of Matter, 22761 Hamburg, Germany}
\affiliation{Department of Physics, Clarendon Laboratory, University of Oxford, Oxford OX1 3PU, United Kingdom}

\author{Ludwig Mathey}
\affiliation{Zentrum f\"ur Optische Quantentechnologien and Institut f\"ur Laserphysik, Universit\"at Hamburg, 22761 Hamburg, Germany}
\affiliation{The Hamburg Centre for Ultrafast Imaging, Luruper Chaussee 149, 22761 Hamburg, Germany}

\date{\today}

\begin{abstract}
Recent pump-probe experiments reported an enhancement of superconducting transport along the $c$ axis of underdoped YBa$_2$Cu$_3$O$_{6+\delta}$ (YBCO), induced by a midinfrared optical pump pulse tuned to a specific lattice vibration. To understand this transient nonequilibrium state, we develop a pump-probe formalism for a stack of Josephson junctions, and we consider the tunneling strengths in the presence of modulation with an ultrashort optical pulse. We demonstrate that a transient enhancement of the Josephson coupling can be obtained for pulsed excitation and that this can be even larger than in a continuously driven steady state. Especially interesting is the conclusion that the effect is largest when the material is parametrically driven at a frequency immediately above the plasma frequency, in agreement with what is found experimentally. For bilayer Josephson junctions, an enhancement similar to that experimentally is predicted below the critical temperature $T_c$. This model reproduces the essential features of the enhancement measured below $T_c$. To reproduce the experimental results above $T_c$, we will explore extensions of this model, such as in-plane and amplitude fluctuations, elsewhere.
%

\end{abstract}

\maketitle
\section{Introduction}

Recent pump-probe experiments  have opened a new field in solid state physics by establishing a method to control material properties via laser pulses in the optical regime.\cite{Forst2015, Mankowsky2016, Nicoletti2016} Several examples are: optical switching of charge-density waves in transition metal dichalcogenides,\cite{Stojchevska2014} creation of effective magnetic fields in rare-earth compounds,\cite{Nova2016} and induction of lattice distortions in manganites.\cite{Rini2007, Tobey2008} In particular, in Refs. \onlinecite{Fausti2011, Kaiser2014, Nicoletti2014, Forst2014, Forst2014a, Hu2014, Mankowsky2014, Hunt2015, Casandruc2015, Mankowsky2015, Khanna2016, Hunt2016, Mankowsky2017, Hu2017}, pump-probe techniques were used to control various layered high-$T_{c}$ superconductors. This resulted in the observations of light-enhanced and light-induced superconductivity. These intriguing experimental results were studied theoretically in Refs. \onlinecite{Denny2015, Raines2015, Hoppner2015, Patel2016, Okamoto2016, Sentef2017}. However, these studies primarily focused on the steady state of this driven system, while the experimental operation uses a pump pulse, with a pulse length that is typically around five times of the inverse optical frequency. It is therefore imperative to study the transient response of the driven system. 

\begin{figure}[!tb]
\begin{center}
\includegraphics[width = 1.0\columnwidth]{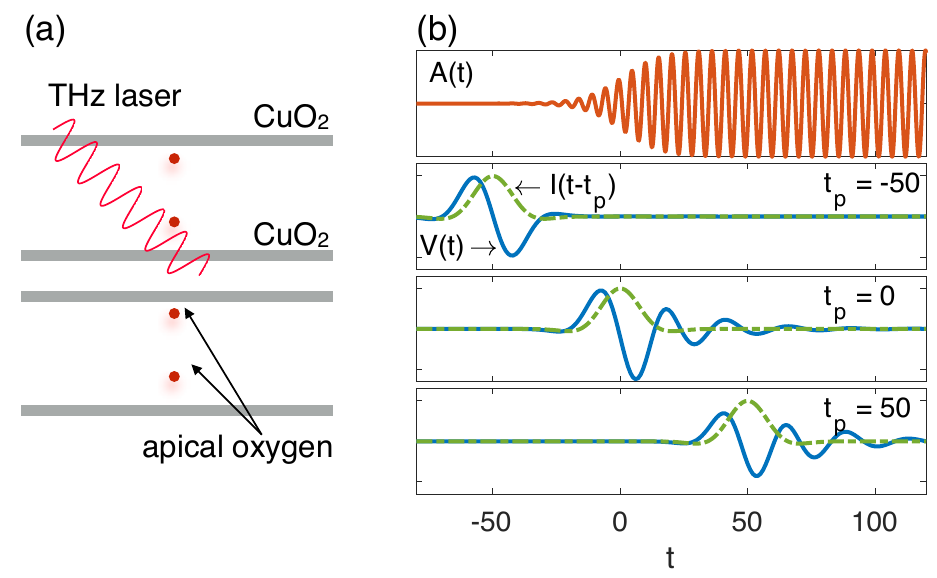}
\caption{(a) Schematic depiction of YBCO. Superconducting CuO$_2$ layers (gray) form a stack of bilayer Josephson junctions. THz pulses (wavy lines) excite apical oxygen atoms (circles)  that induce oscillations of $j_{1}$ and $j_{2}$. (b) Typical time-dependent voltage response $V(t)$ (solid lines) for different probe pulses $I(t-t_p)$ (dashed lines) at $t_p = -50, 0$ and $50$. The driving amplitude $A(t)$ is also depicted.}
\label{geometry}
\end{center}
\end{figure}

In this paper, we study the transient response of the superconducting phase below the critical temperature $T_c$ in layered systems, which we model as capacitively coupled Josephson junctions (see Fig.~\ref{geometry}).\cite{VanderMarel1996, Koyama1996, Matsumoto1999, Koyama1999, Machida1999, Koyama2000, VanderMarel2001, Koyama2001, Koyama2002, Machida2004, Machida2004a, Shukrinov2007,Shukrinov2009} The model is limited by its low dimensionality and lack of amplitude fluctuations of the order parameter, which prohibits us to describe light-induced superconductivity far above $T_c$. steady state properties of similar models have been investigated in Refs.~\onlinecite{Denny2015, Hoppner2015, Okamoto2016}. Here, in order to obtain the time-resolved conductivity, we introduce a pump-probe scheme similar to the one used  experimentally by scanning through various pump-probe delay times with narrow probe pulses. In Sec. \ref{singleA}, we first consider a single Josephson junction as a simple model for the interlayer phase dynamics. When the frequency of the parametric driving is just above the Josephson plasma frequency, the effective Josephson coupling both in the transient and the driven steady state is increased. In particular, when the driving pulse is narrow in time, the transient value can be larger than the steady state value.  We also find that an effective critical temperature $T_c$ of the transient state, as defined below, can be larger than that of the steady state. In Sec. \ref{singleB}, we first relate the transient behavior to driving the junction with additional higher harmonic frequencies. We then extend this analysis to point out an improved driving method that combines several harmonics in steady states. In Sec. \ref{Bilayer}, we use an effective model of a stack of weak and strong junctions, resembling the structure of YBCO.\cite{VanderMarel1996, VanderMarel2001, Koyama2002} Again, we find a transient enhancement of the Josephson coupling, and the comparison with experimental data shows qualitative agreement below $T_c$. Better quantitative description of the light-enhanced and -induced superconductivity needs to go beyond our model and include more complex physics such as amplitude fluctuations, lattice distortions, and competing charge order. Finally, Sec. \ref{conclusion} is the conclusion.

\section{Single Josephson junction: Transient dynamics}\label{singleA}
\subsection{Model and Method}
As our first model, we study a single Josephson junction with a bare Josephson coupling $J_0$, a thickness $d$, and a dielectric constant $\epsilon$. It has a characteristic plasma frequency $\omega_\text{Jp} = \sqrt{4\pi e^* d J_0 / \hbar \epsilon}$. The phase $\varphi$ of the junction obeys
\begin{equation}
\ddot{\varphi} + \gamma \dot{\varphi} + \omega_\text{Jp}^2 \left[1 + A(t, t_e) \right] \sin \varphi = I + \xi,
\label{single_SG0}
\end{equation}
where $\gamma$ is a damping coefficient, $I$ an external current, and $\xi$ the thermal noise characterized by a temperature $T$ via $\langle \xi(t) \xi(t')\rangle = 2 \gamma k_B T \delta(t-t')$. We have included a parametric modulation of $J_0$ with an amplitude $A$ as $J_0 \rightarrow  J_0 \left[1 + A(t, t_e) \right]$.\cite{Jung1993, Zerbe1994, maclachlan1964theory, nayfeh2008nonlinear, Citro2015, Zhu2016} As we will discuss in more detail later, modulation of $J_0$ may be induced by optically excited oxygen atoms inside the junction. Mathematically, the result does not change if the dielectric function $\epsilon$ or the interlayer thickness $d$ is modulated; they all periodically change $\omega_\text{Jp}$ and drive the junction parametrically. As the pump or driving pulse $A(t, t_e)$, we choose either a continuous driving pulse with a nonzero rise time
\begin{equation}
A(t, t_e) =  \frac{A_{0}}{2} \cos(\omega_e t + \phi) \left[ \tanh \left(\frac{t-t_e}{\Delta_{e}} \right) +1\right]
\label{A2}
\end{equation}
or a Gaussian pulse 
\begin{equation}
A(t, t_e) = A_{0} \cos(\omega_e t + \phi) \exp \left[- \frac{(t-t_{e})^{2}}{2 \Delta_{e}^{2}} \right].
\label{A1}
\end{equation}
For both, $A_{0}$ is the amplitude of the driving, $\omega_e$ the driving frequency, $\phi$ the initial phase, and $\Delta_{e}$ the rise time or the pulse length, respectively. $t_e$ characterizes the starting time of the driving. The continuous driving gives access to the relaxation to the steady state, while the pulsed driving can illuminate short transient dynamics. We assume that the phase $\phi$ is uncontrolled, which is the case for the experiments discussed here.\cite{Jung1993, Zerbe1994} In the following we always take a phase average over $\phi \in [0, 2\pi]$.

\begin{figure}[!tb]
\begin{center}
\includegraphics[width = \columnwidth]{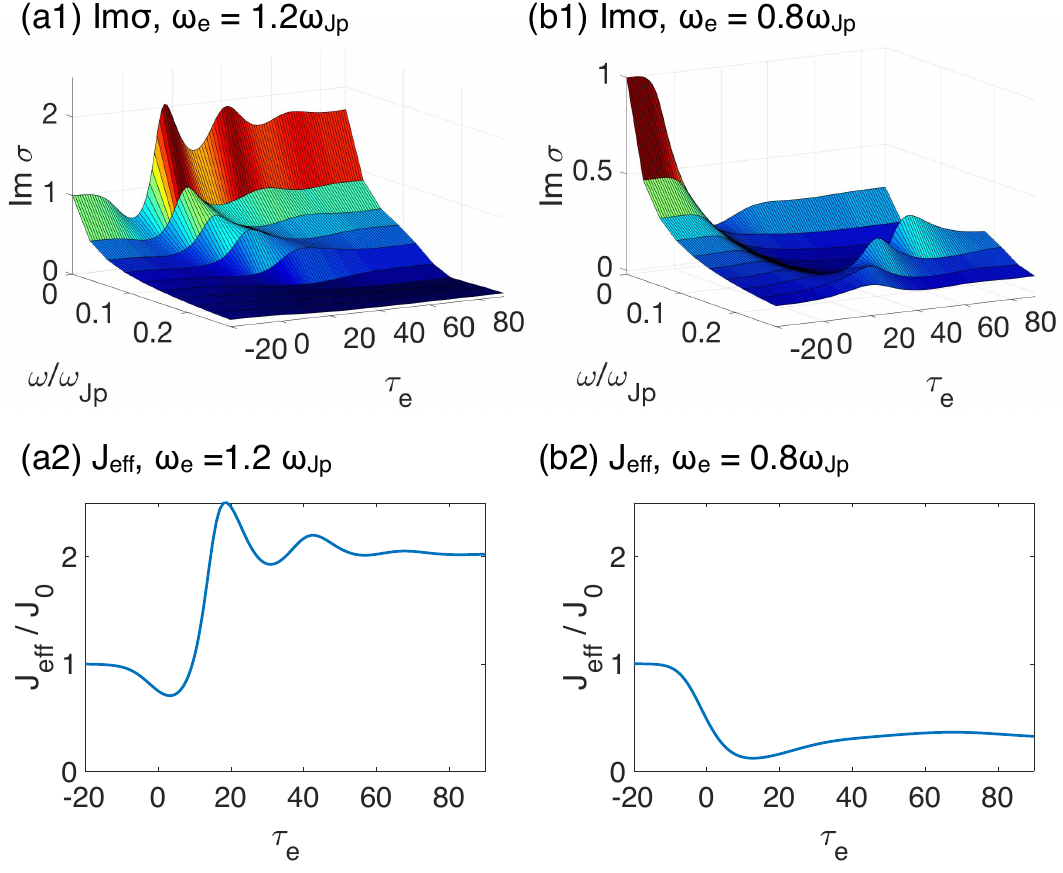}
\caption{Transient imaginary conductivity $\Im \sigma (\omega, \tau_e)$ and $J_\text{eff}(\tau_e)$ for continuous driving at $T=0$. (a) Blue-detuned case $\omega_e = 1.2 \omega_\text{Jp}$. (b) Red-detuned case $\omega_e = 0.8 \omega_\text{Jp}$.}
\label{Tanh_sigma}
\end{center}
\end{figure} 
 
In order to obtain a time-resolved conductivity, we follow the formulation of Ref. \onlinecite{Kindt1999} (see also Refs. \onlinecite{Nemec2002, Orenstein2015, Shao2016, Kennes2017b}.) We add a probe pulse to the system, 
\begin{equation}
I(t- t_p) = I_{0} \cos\left[\omega_p (t-t_p) \right] \exp \left[- \frac{(t-t_{p})^{2}}{2 \Delta_{p}^{2}} \right],
\end{equation}
and then measure the voltage $V$ across the junction at sampling time $t_s$. We fix the pump time $t_e$ and scan $t_p$ and $t_s$. The number of probe pulses during a fixed time window and the shape of the spectrum determines the resolution of the obtained data. Without a driving pulse, the response of the system depends only on the difference $t_s - t_p$. However, with the time-dependent driving pulse, this is no longer the case (see Fig. \ref{geometry}), and the resistivity response $\rho$ becomes time dependent
\begin{equation}
V(t_s-t_p, t_s - t_e) = \int_{-\infty}^{t_s} \rho(t_s - t', t_s - t_e) I(t'- t_p) dt' .
\end{equation}
Moving to the relative time variables $\tau \equiv t_s-t_p$ and $\tau_e \equiv t_s - t_e$, we rewrite this as a convolution,
\begin{equation}
V(\tau, \tau_e) = \int_{-\infty}^{\tau} \rho(\tau - t', \tau_e) I(t') dt' .
\label{convol}
\end{equation}
Fourier transforming the above equation in terms of $\tau$, we define the time-dependent conductivity as
\begin{equation}
\sigma(\omega, \tau_e) \equiv \frac{1}{\rho(\omega, \tau_e)} = \frac{I(\omega) d}{V(\omega, \tau_e)} .
\label{sigma_def}
\end{equation}
This quantity resembles the transient conductivity that was measured in Ref. \onlinecite{Hu2014}. As in Ref. \onlinecite{Okamoto2016} we define an effective Josephson coupling $J_\text{eff}$ via
\begin{equation}
J_\text{eff} (\tau_e) \equiv \frac{\hbar}{e^* d} \Im[\sigma(\omega, \tau_e) \omega]_{\omega \rightarrow 0}.
\label{J_eff}
\end{equation}
This reduces to $J_0$ in equilibrium, and thus quantifies the effective interlayer tunneling energy.

\subsection{Transient conductivity}
In Fig. \ref{Tanh_sigma}, we first show $\Im \sigma (\omega, \tau_{e})$ and $J_\text{eff}$ for continuous driving with $A_0 = 0.8$, $\Delta_e = 15$, and $\gamma = 0.1$ at $T=0$ (in the following, we put $\omega_\text{Jp}  = 1$). The probe pulse is taken as $I_0 = 0.1$, $\omega_p = 0.1 \omega_\text{Jp}$, and $\Delta_p = 10$. We numerically integrate the equation of motion by Heun scheme with time step $h = 10^{-3}$. As we have shown in Ref. \onlinecite{Okamoto2016}, the interlayer tunneling is enhanced (suppressed) at the blue- (red-) detuned side, and the driven steady state value is approximately  
 \begin{equation}
J_\text{eff}^\text{steady} \simeq J_0\left[ 1 + \frac{A_0^2 \omega_\text{Jp}^2 (\omega_e^2 - \omega_\text{Jp}^2 )}{2(\omega_e^2 - \omega_\text{Jp}^2)^2 +2 \gamma^2 \omega_e^2}\right] .
\label{Jeff_steady}
 \end{equation}
Interestingly, for $\omega_{e} > \omega_\text{Jp}$, the transient value of $J_\text{eff}$ first shows a dip in the initial stage of the driving, followed by a large peak, and then reaches to the steady state after a few small oscillations. We will explain this behavior in more detail below.

\begin{figure}[!tb]
\begin{center}
\includegraphics[width = \columnwidth]{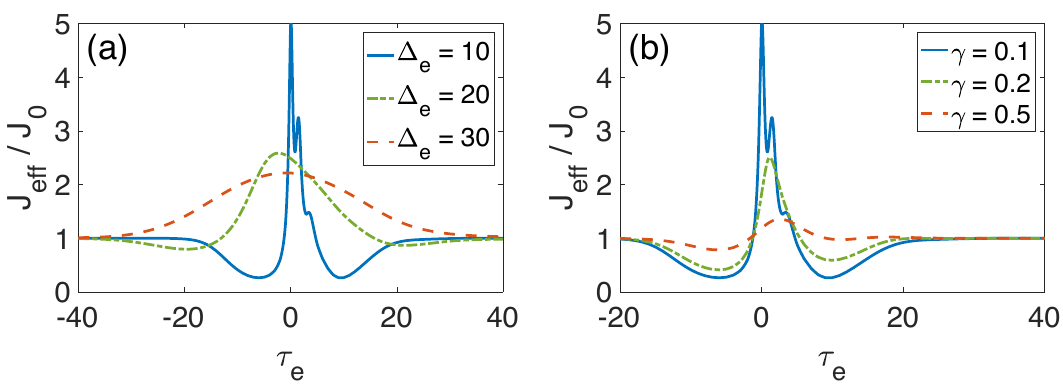}
\caption{(a) $J_\text{eff} (\tau_e)$ for several pump widths $\Delta_e$ for $\gamma = 0.1$ at $T=0$. (b) $J_\text{eff} (\tau_e)$ for different damping factors $\gamma$ for $\Delta_e = 10$ at $T=0$.}
\label{Gaussian_Jeff}
\end{center}
\end{figure}

To  elucidate the transient behavior further, we consider a Gaussian pulse, Eq.~\eqref{A1}, with $A_0= 0.8$ and $\omega_e = 1.2 \omega_\text{Jp}$. We plot $J_\text{eff}$ for  several pump durations in Fig. \ref{Gaussian_Jeff}(a) with $\gamma = 0.1$. We find that the transient peak around $\tau_e = 0$ is larger for smaller pump duration and that it is accompanied by a dip before and after the peak. When the duration is long enough, e.g., $\Delta_e = 30$, we  observe the enhancement of $J_\text{eff}$ only. Figure~\ref{Gaussian_Jeff}(b) shows the damping dependence of $J_\text{eff}$ at $\Delta_e = 10$. When the damping is increased, the peak value decreases significantly, and at the same time the dip diminishes.

\subsection{Temperature dependence}
\begin{figure}[!tb]
\begin{center}
\includegraphics[width = \columnwidth]{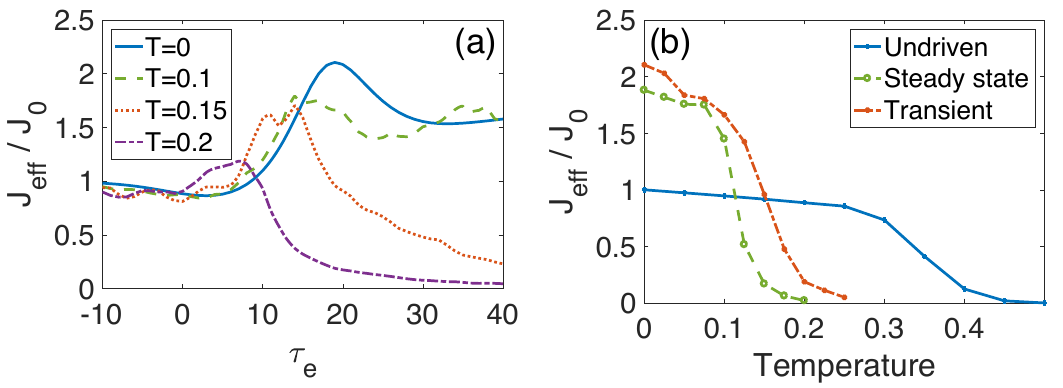}
\caption{(a) Temperature dependence of $J_\text{eff} (\tau_e)$ for continuous driving. We use $\gamma = 0.1$, $\Delta_e = 15$, and $\omega_e = 1.2$. (b) Temperature dependence of $J_\text{eff}$ for $A_0= 0$ (undriven), $A_0=0.8$ (steady state), and $A_0=0.8$ (transient value at $\tau_e = 19$).}
\label{temperature}
\end{center}
\end{figure}

Figure \ref{temperature}(a) illustrates the temperature dependence of $J_\text{eff}$ for continuous driving starting at $\tau_e  =0$. The parameters of the model are the same as the blue-detuned side of Fig. \ref{Tanh_sigma}, and the results are  averaged over $\sim 10^5$ samples. At low temperatures, after a few oscillations, the system approaches the driven steady state with  a nonzero $J_\text{eff}$. As the temperature increases, while the peak still appears, it is shifted to earlier times. The following steady state has vanishingly small Josephson coupling (c.f. the purple line for $T=0.2$); the transient state has higher $T_c$ than the steady state. In Fig. \ref{temperature}(b), we compare $J_\text{eff}(T)$ of the undriven, steady state, and the transient (at $\tau_e = 19$) values. We observe that the driven state has a lower $T_c$ compared to the undriven one. As we have shown in Ref.~\onlinecite{Okamoto2016}, this is due to the fact that the fluctuations of the current $j(t) \equiv \sin \varphi(t)$ are increased  when integrated over all frequencies $\int d\omega \langle |j(\omega)|^2 \rangle$. This increased noise induces phase slips and destroys the phase coherence, even though the low-frequency part of the fluctuations, which determines $J_\text{eff}$, is reduced. \footnote{Importance of phase slips was discussed recently in Ref. \onlinecite{Kennes2017a}. } A similar reduction of $T_c$ is seen in the bilayer case, when the plasma frequency of the weak junction is driven directly. \cite{Okamoto2016} 

\section{Single Josephson junction: higher-order harmonic driving}
\label{singleB}
In the previous section, we have demonstrated that the transient response of a single Josephson junction deviates from the steady state response under monochromatic driving. Here, we attribute the deviation to mixing of higher order driving frequencies, in particular  $\pm n\omega_e$ with $n = 2, 3, \cdots$, that appear in the driving pulse. To elaborate on this observation, we discuss the response function of a Josephson junction that is driven by two frequencies. We present both a numerical calculation and an analytical estimate. These results expand on the response function that was derived in Ref.~\onlinecite{Okamoto2016} for a single driving frequency. We assume that the driving pulse has an admixture of the second harmonic frequency,
\begin{equation}
\begin{split}
A(t) &= A_0 \cos (\omega_e t ) + A_1 \exp(i2 \omega_e t) + A_1^* \exp(-i2 \omega_e t)\\
&= A_0 \cos (\omega_e t ) + |A_1| \cos(2 \omega_e t + \phi_1),
\end{split}
\end{equation}
with $\omega_e$ being the principal driving frequency. $A_1 = |A_1| e^{i \phi_1}$ is the complex amplitude for the second order harmonic driving, where $\phi_1$ gives the phase difference between the first harmonic and the second one. By Fourier transforming and linearizing Eq. \eqref{single_SG0}, we obtain
\begin{multline}
\left(\frac{-\omega^2 - i \gamma \omega}{\omega_\text{Jp}^2} + 1 \right) \varphi(\omega) + \frac{A_0}{2} \left[ \varphi (\omega + \omega_e) + \varphi (\omega - \omega_e)\right] \\
+ \frac{1}{2} \left[ A_1 \varphi (\omega + 2\omega_e) + A_1^*\varphi (\omega - 2\omega_e) \right] =  \frac{I (\omega)}{\omega_\text{Jp}^2}.
\end{multline}
We assume that the external current is monochromatic $I(\omega) = I_0 \delta(\omega - \omega_p)$ and the probing frequency $\omega_p$ is taken to be much smaller than the driving frequency $\omega_e$ and the plasma frequency $\omega_\text{Jp}$, since we are interested in the low frequency conductivity. This allows us to write down a discrete set of coupled equations for $\varphi_n \equiv \varphi (\omega_p + n \omega_e)$ ($n \in \mathbb{Z}$) as 
\begin{equation}
\frac{1}{2}
\begin{bmatrix}
\ddots &&&&& \\
 2 K_2  & A_0 & A_1^* & 0 & 0  \\
 A_0 & 2 K_1 & A_0 & A_1^* & 0  \\
 A_1 & A_0 & 2 K_0 & A_0 & A_1^*  \\
 0 & A_1 & A_0 & 2 K_{-1} & A_0  \\
 0 & 0 & A_1 & A_0 &  2K_{-2}  \\
&&&& \ddots
\end{bmatrix}
\begin{bmatrix}
\vdots \\
\varphi_2 \\
\varphi_1 \\
\varphi_0 \\
\varphi_{-1} \\
\varphi_{-2} \\
\vdots
\end{bmatrix}
=
\begin{bmatrix}
\vdots \\
0 \\
0 \\
\frac{I_0}{\omega_\text{Jp}^2} \\
0 \\
0 \\
\vdots
\end{bmatrix},
\label{S2}
\end{equation}
where the diagonal elements are given by
\begin{equation}
K_n \equiv \frac{-\left(\omega_p+n \omega_e\right)^2 - i \gamma (\omega_p+ n \omega_e )}{\omega_\text{Jp}^2} + 1 .
\end{equation}
\begin{figure}[!tb]
\begin{center}
\includegraphics[width = 0.9\columnwidth]{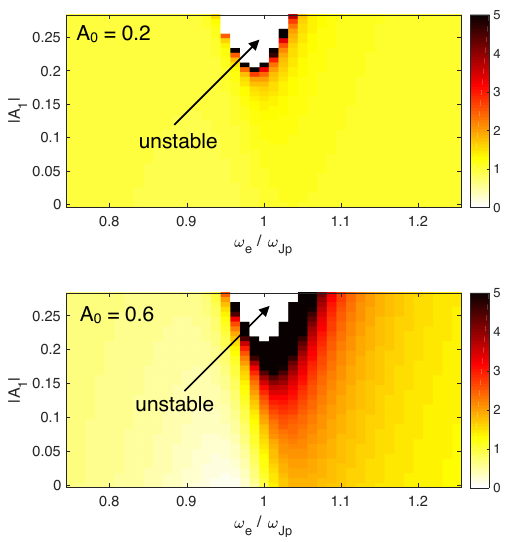}
\caption{$J_\text{eff}/J_0$ in the driven steady state with additional second harmonic driving $|A_1| \cos(2\omega_e t + \phi_1)$ as a function of $|A_1|$ and $\omega_e$ at $\gamma =0.1$, and $A_0 = 0.2$ (top) or $0.6$ (bottom). The phase difference $\phi_1$ is taken to be 0.}
\label{SM1}
\end{center}
\end{figure}

We obtain the solution by inverting the matrix numerically. We truncate the infinite matrix equation by taking $21$ modes ($n = -10, \dots, 10$); we have checked that the convergence is well achieved in terms of the number of modes. Once we find $\varphi_{0} = \varphi(\omega_p)$, we compute the conductivity from the Josephson relation $V = (\hbar/e^*) \dot{\varphi}$ and $\sigma = I d/ V$ as
\begin{equation}
\sigma(\omega_p) =  i \left( \frac{e^* d}{\hbar } \right) \frac{I_0}{\omega_p \varphi (\omega_p)}.
\end{equation}
We define the effective Josephson coupling as in Eq. \eqref{J_eff}. 

First, we discuss the case of in-phase driving, i.e., $\phi_1 = 0$. In Fig.~\ref{SM1}, we plot $J_\text{eff}$ as a function of $|A_1|$ and $\omega_e$ for $A_0 = 0.2$, and $0.6$ at $\gamma = 0.1$.  We have excluded the regions where the driving pulse leads to Floquet parametric instability. \cite{coddington1997linear, cesari2013asymptotic} The stability is determined by the Floquet exponent obtained by integrating the equation of motion for one period of the driving, $2\pi/\omega_e$, with initial conditions $\left[\varphi (0), \dot{\varphi} (0) \right] = [1, 0]$ and $[0,1]$.\cite{coddington1997linear, cesari2013asymptotic} We note that the instability region depends only weakly on the driving amplitude of the primal harmonic $A_0$. This indicates that the instability mainly comes from the second harmonic driving. Remarkably, the weak additional harmonic $A_1$ gives rise to larger values of $J_\text{eff}$ for the blue detuned side. As a competing effect, for larger values of $A_1$, the system reaches the primary Floquet instability lobe.

\begin{figure}[!tb]
\begin{center}
\includegraphics[width = \columnwidth]{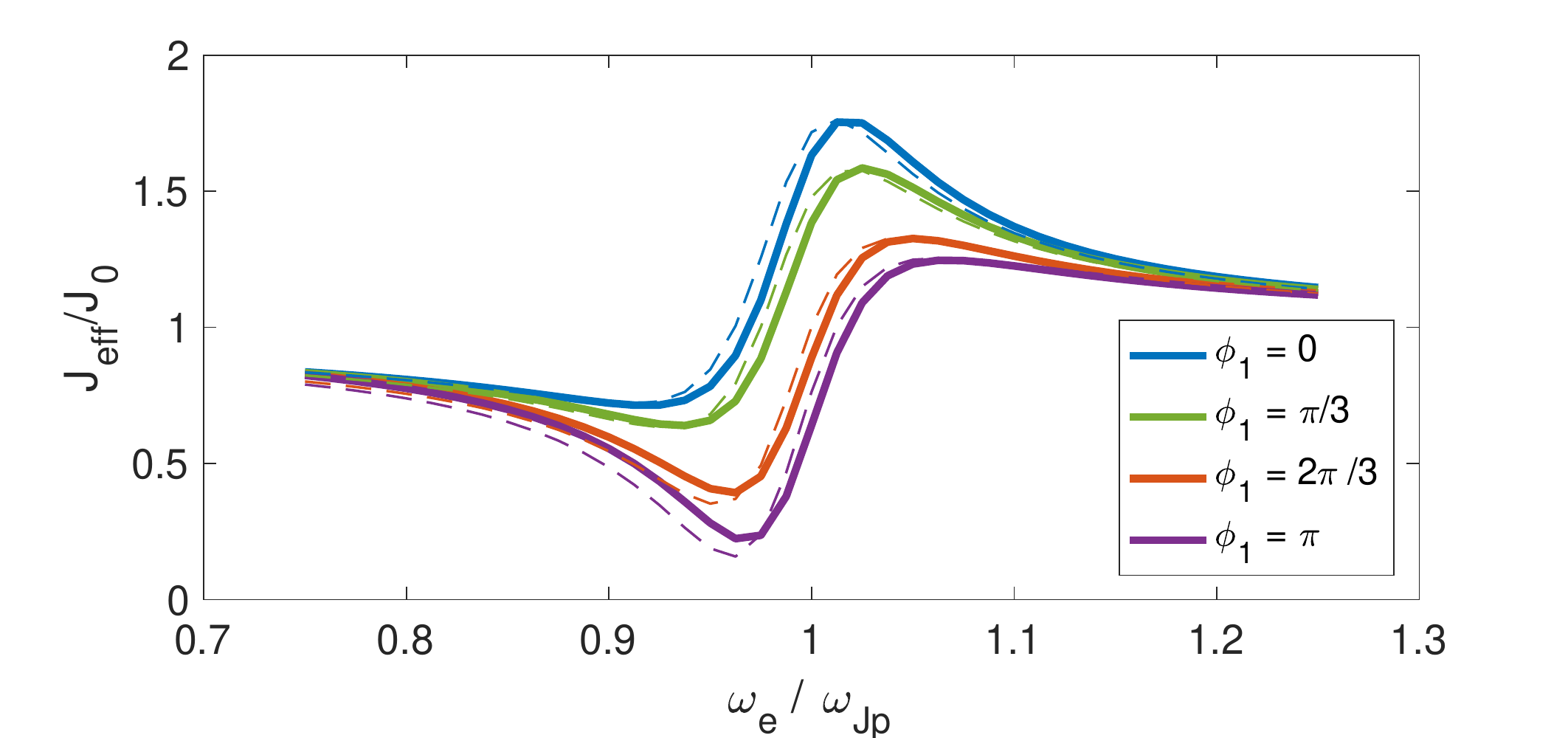}
\caption{The effective Josephson energy $J_\text{eff}$ for various phase differences $\phi_1$ at $A_0=0.4$, $|A_1| = 0.1$, and $\gamma=0.1$. The solid lines are obtained by taking 21 modes, and the dashed line by 3 modes, Eq.~\eqref{Jeff_2nd}.}
\label{SM2}
\end{center}
\end{figure}

This result is modified by different values of the phase $\phi_1$. Figure~\ref{SM2} shows $J_\text{eff}$ as a function of $\omega_e$ at $A_0 = 0.4$, $|A_1| = 0.1$, and $\gamma = 0.1$ for various phase differences $\phi_1$. We find that the largest enhancement of $J_\text{eff}$ is obtained for in-phase driving $\phi_1= 0$. As we dephase the two driving, the enhancement becomes weaker monotonically. To better understand this result, let us consider Eq.~ \eqref{S2} with only three modes ($n=-1, 0, 1$). The analytical expression of $J_\text{eff}$ is obtained as
\begin{equation}
\frac{J_\text{eff}}{J_0} \simeq  1+ A_0^2 \omega_\text{Jp}^2 \frac{2 \omega_e^2 +(|A_1| \cos \phi_1 -2 )\omega_\text{Jp}^2 }{4(\omega_e^2 - \omega_\text{Jp}^2)^2 - |A_1|^2 \omega_\text{Jp}^4  + 4 \gamma^2 \omega_e^2 },
\label{Jeff_2nd}
\end{equation}
which is a generalization of Eq.~\eqref{Jeff_steady}. The second term is the correction by the driving. For $\omega_e > \omega_\text{Jp}$, the numerator becomes bigger when the two driving are near in-phase regime, $ 0 \leq \phi_1  < \pi/2$, while the near out-of-phase driving, $ \pi/2 \leq \phi_1  \leq \pi$, leads to a smaller numerator. Overall tendency thus depends on the phase difference $\phi_1$. For small $A_1$, the numerator dominantly decides if $J_\text{eff}$ is enhanced or reduced compared to the $A_1=0$ case, since the denominator has only the quadratic contribution of $A_1$. 

We note that these results suggest that $J_\text{eff}$ can be maximized in the driven steady state by carefully designing multifrequency optical driving, which takes advantage of the increase that can be achieved by adding higher harmonics, while avoiding the parametric instability regime. These two features also compete in the transient response due to a short driving  pulse. In particular, at the initial stage of a short pulse, e.g., $\Delta_e=10$, the system is effectively driven by higher harmonics, in addition to the base frequency, which can lead to an initial  suppression of $J_\text{eff}$, and then to strong increase of  $J_\text{eff}$, as the higher harmonic admixture is reduced in time, crossing through the regime of optimal admixture.

\section{Bilayer Josephson junctions}\label{Bilayer}
As our second model, we consider a stack of alternating weak ($i=1$) and strong ($i=2$) junctions (see Fig.~\ref{geometry}). Each junction is characterized by a thickness $d_{i}$, a dielectric constant $\epsilon_{i}$, a Josephson critical current $j_{i}$, and a bare plasma frequency $\Omega_{i} = \sqrt{4\pi e^* d_{i} j_{i} / \hbar \epsilon_{i}}$. We ignore fluctuations among different unit cells. Then the equation of motion of the phase differences $\varphi_i$ becomes,\cite{VanderMarel1996, VanderMarel2001, Koyama2002, Okamoto2016} 
\begin{multline}
\begin{bmatrix}
\ddot{\varphi_1} \\
\ddot{\varphi_2}
\end{bmatrix}
+ \gamma 
\begin{bmatrix}
\dot{\varphi_1} \\
\dot{\varphi_2}
\end{bmatrix}
- \frac{4\pi e^* \mu^2 I}{ s}
\begin{bmatrix}
\alpha^{-1}_1 \\
\alpha^{-1}_2
\end{bmatrix}\\
 = 
 \begin{bmatrix}
-(1 + 2\alpha_1)\Omega_1 ^2& 2\alpha_2 \Omega_2 ^2 \\
 2\alpha_1 \Omega_1 ^2  & -(1 + 2\alpha_2)\Omega_2 ^2 
 \end{bmatrix}
 \begin{bmatrix}
\varphi_1 \\
\varphi_2
\end{bmatrix}.
\label{SG_eq0}
\end{multline}
 $\alpha_i \equiv \epsilon_i \mu^2 / s d_i$ is the capacitive coupling constant with $s$ being the thickness of the superconducting layer and $\mu$ the Thomas-Fermi screening length in the superconducting layers. The voltage is related to the phase differences by the generalized Josephson relations,\cite{Koyama1996, Koyama2002}
\begin{equation}
\left( \frac{\hbar}{e^*}\right)
\begin{bmatrix}
\dot{\varphi_1} \\
\dot{\varphi_2}
\end{bmatrix}
 = 
 \begin{bmatrix}
1 + 2\alpha_1& -2\alpha_2 \\
 -2\alpha_1  & 1 + 2\alpha_2 
 \end{bmatrix}
 \begin{bmatrix}
V_1 \\
V_2
\end{bmatrix}.
\label{general_Josephson}
\end{equation}
For the undriven case at $T=0$, Eqs.~\eqref{SG_eq0} and \eqref{general_Josephson} give
\begin{equation}
\sigma (\omega ) = \frac{\epsilon_\text{av}}{4\pi i} \frac{\left( \omega^2 + i \gamma \omega - \omega_\text{Jp1}^2 \right) \left( \omega^2 + i \gamma \omega- \omega_\text{Jp2}^2 \right)}{ \omega \left( \omega^2 + i \gamma \omega - \omega_{t}^2 \right)},
\label{conductivity}
\end{equation}
where $\omega_\text{Jp1, Jp2} \simeq \Omega_{1,2}$ are the longitudinal plasma modes for weak and strong junctions, and $\omega_t \simeq \omega_\text{Jp2}$ is the transverse plasma mode.\cite{Koyama2002} We take the parameters of the model as $\alpha_1 = 3$, $\alpha_2 = 1.5$, $\Omega_1 = 1$, $\Omega_2 = 12.5$, and $\gamma = 0.1$. These are chosen to reproduce the ratio $\omega_\text{Jp2}/\omega_\text{Jp1}\sim 15.8$ of YBCO with appropriate $\alpha$ values for this compound of around $\sim 3$.\cite{Koyama2002} We have  $\omega_\text{Jp1} = 1.58$ and $\omega_\text{Jp2} = 25.1$. The probing pulse is taken as $I_0 = 0.1$, $\omega_p = 1.5$, and $\Delta_p = 1$ so that the frequencies around $\omega_\text{Jp1}$ are well resolved as the experimental condition.

\begin{figure}[!tb]
\begin{center}
\includegraphics[width = 0.8\columnwidth]{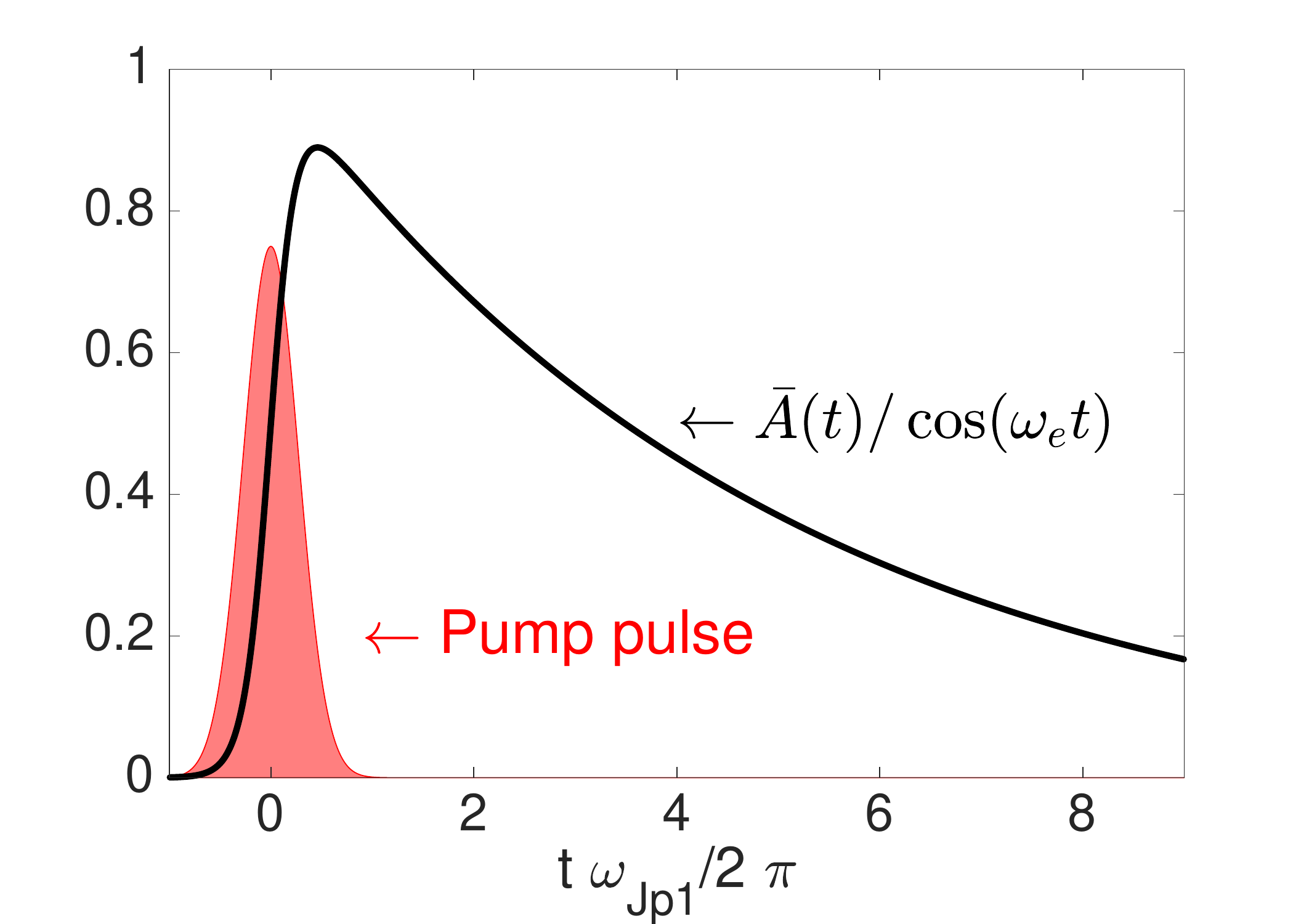}
\caption{Schematic description of the relationship between the optical pump pulse that excites the apical oxygen atoms and the driving amplitude given in Eq.~\eqref{Abar} (only the envelop parts are depicted).}
\label{pump_driving}
\end{center}
\end{figure}

In the experiment by Hu \textit{et al.},\cite{Hu2014} the optical pump pulse has a period of $\sim 50$fs and its duration is $\sim 0.3$ps, while the lifetime of the resonantly driven infrared $B_{1u}$ mode, which displaces apical oxygens along the $c$ axis, exceeds $2$ps.\cite{Subedi2014, Mankowsky2014, Mankowsky2015} The oxygen motion primarily affects the interlayer motion of Cooper pairs, and thus we assume that the modulation of the Josephson critical currents $j_i$ derives from this driven phonon mode. The modulation of $j_i$ leads to parametric driving of the Josephson junction. As in the single Josephson junction case, we note that such parametric driving may be realized by other mechanisms such as modulation of the dielectric function,\cite{Denny2015}. However, the microscopic origin of the modulation is not important for the following discussion. To imitate the transient phonon motions, we take the driving as
\begin{multline}
\bar{A}(t) =  \frac{1}{2} \cos(\omega_e t +\phi)  \left[ \tanh \left(\frac{t}{\Delta_{e}} \right) +1\right] e^{- \gamma_e t},
\label{Abar}
\end{multline}
with $\omega_e = 26$, $\Delta_e = 1$, and $\gamma_e = 0.05$. This shows a sharp rise within several cycles of $\omega_e$ and then an exponential decay over dozens of cycles (Fig.~\ref{pump_driving}). This parametric driving is included by changing the critical currents as
 $j_{1,2} \rightarrow j_{1,2} [1\pm a_{1,2} \bar{A}(t)]$; we assume that the driving is alternating along the junctions.
 
\begin{figure}[!tb]
\begin{center}
\includegraphics[width = \columnwidth]{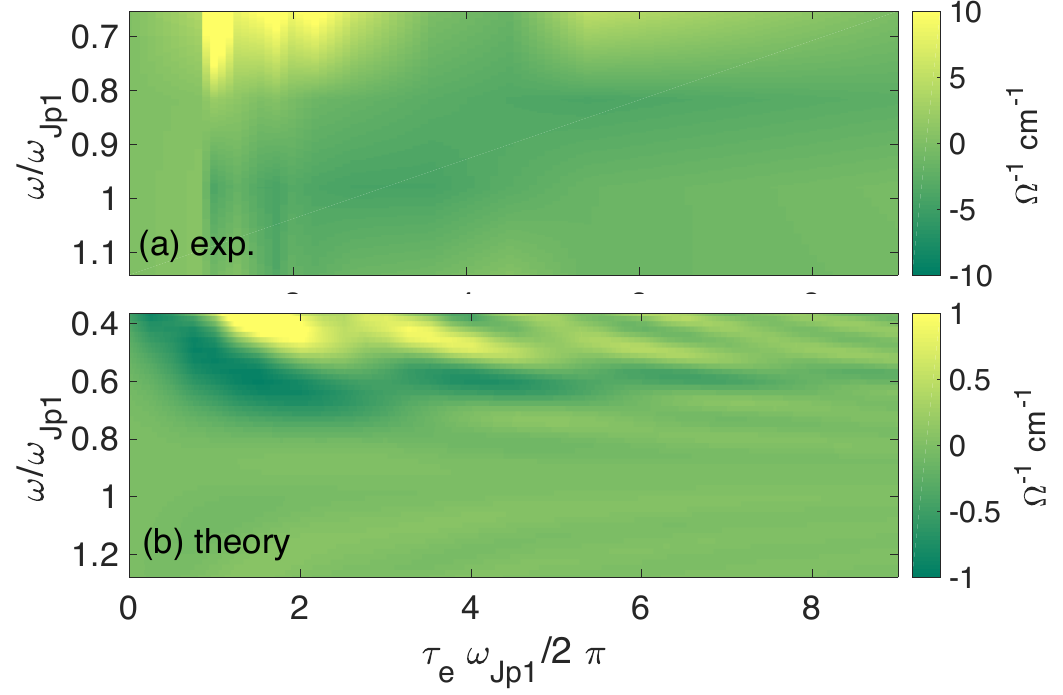}
\caption{Changes in transient imaginary conductivity $  \Im \Delta \sigma(\omega, \tau_e)$. The simulated conductivity is rescaled to fit the value at $\omega = 0.5 \omega_\text{Jp1}$ in equilibrium to the experimental one. (a) Experimental data from Ref.~\onlinecite{Hu2014} at $T =10\ \text{K} \ll T_c$. (b) A simulated result at $\gamma = 0.1$, $a_1 = 0.3$, and $a_2 = 0.6$ at $T=0$. }
\label{bilayer}
\end{center}
\end{figure}

In Fig.~\ref{bilayer}, we compare the change of conductivity $\Im \Delta \sigma (\omega, \tau_e) \equiv \Im \sigma (\omega, \tau_e) - \Im \sigma^\text{eq}(\omega)$ obtained by simulations to the experimental result of Ref.~\onlinecite{Hu2014} at $T =10\text{K} \ll T_c$. For the simulation, we take $a_1 = 0.3$, $a_2 = 0.6$, and $T= 0$. At low frequencies, on the rise of the driving, a peak appears after a small dip in the simulation. This is similar to the single junction case. The peak is followed by a decay over few oscillations, relaxing back to the original state. The period of such oscillations is approximately one cycle of $\omega_\text{Jp1}$. We also observe that the transient change $\Im \Delta \sigma(\omega, \tau_e)$ becomes negative at high frequencies in the simulation. This overall transient behavior of the simulation is qualitatively similar to the experimental one, while a few discrepancies remain. For instance, a dip in the initial stage of the driving and subsequent small oscillations are absent in the experiment. Also, the relative enhancement at low frequencies $\sim 0.5 \omega_\text{Jp1}$ in the simulation is $\sim 15 \%$ at most, while that in the experiment is $\sim 100 \%$.\footnote{In equilibrium, $\Im\sigma(\omega= 0.5 \omega_\text{Jp1}) \approx 10 \ \Omega^{-1} \text{cm}^{-1}$.} These discrepancies may arise due to physics that is not included in our simulation such as finite temperature effects, amplitude fluctuations of the order parameter, nonlinear lattice distortion,\cite{Mankowsky2014, Raines2015} and competing orders.\cite{Patel2016, Sentef2017}

\section{Conclusions}\label{conclusion}
In this paper, we have studied transient superconductivity in optically driven high $T_c$ superconductors using Josephson junction models below $T_c$. We find that the transient state shows enhanced interlayer tunneling, which can be larger than the steady state value, when the system is driven near the blue-detuned side of the Josephson plasma frequency. We have explained the transient behavior by considering the higher order harmonics in driving.  We have also shown that our bilayer model can phenomenologically explain the temporal change of the imaginary part of conductivity seen in experiments on YBCO below $T_c$, while quantitative differences still remain; in particular it can hardly explain the light-induced superconductivity. The differences may derive from more complex physics including amplitude fluctuations, lattice distortion,\cite{Mankowsky2014, Raines2015} or competing charge order.\cite{Patel2016, Sentef2017} We have also demonstrated that admixing higher harmonics in the driving operation can result in an additional enhancement of the $c$-axis transport. This observation opens the door towards optimal control of superconductivity via optical driving, by combining several higher harmonics. It is an interesting open question if the conductivity of phonon driven BCS superconductors \cite{Mitrano2016, Knap2016, Kim2016, Komnik2016, Sentef2016, Kennes2017, Murakami2017, Mazza2017, Babadi2017, Nava2017} also shows larger enhancement when the higher-order harmonic driving is mixed.

\acknowledgements

We acknowledge the support from the Deutsche Forschungsgemeinschaft (through SFB 925 and EXC 1074) and from the Landesexzellenzinitiative Hamburg, which is supported by the Joachim Herz Stiftung. J.O. thanks S. A. Sato for helpful discussions about details of the pump-probe formalism.


\begin{thebibliography}{67}%
\makeatletter
\providecommand \@ifxundefined [1]{%
 \@ifx{#1\undefined}
}%
\providecommand \@ifnum [1]{%
 \ifnum #1\expandafter \@firstoftwo
 \else \expandafter \@secondoftwo
 \fi
}%
\providecommand \@ifx [1]{%
 \ifx #1\expandafter \@firstoftwo
 \else \expandafter \@secondoftwo
 \fi
}%
\providecommand \natexlab [1]{#1}%
\providecommand \enquote  [1]{``#1''}%
\providecommand \bibnamefont  [1]{#1}%
\providecommand \bibfnamefont [1]{#1}%
\providecommand \citenamefont [1]{#1}%
\providecommand \href@noop [0]{\@secondoftwo}%
\providecommand \href [0]{\begingroup \@sanitize@url \@href}%
\providecommand \@href[1]{\@@startlink{#1}\@@href}%
\providecommand \@@href[1]{\endgroup#1\@@endlink}%
\providecommand \@sanitize@url [0]{\catcode `\\12\catcode `\$12\catcode
  `\&12\catcode `\#12\catcode `\^12\catcode `\_12\catcode `\%12\relax}%
\providecommand \@@startlink[1]{}%
\providecommand \@@endlink[0]{}%
\providecommand \url  [0]{\begingroup\@sanitize@url \@url }%
\providecommand \@url [1]{\endgroup\@href {#1}{\urlprefix }}%
\providecommand \urlprefix  [0]{URL }%
\providecommand \Eprint [0]{\href }%
\providecommand \doibase [0]{http://dx.doi.org/}%
\providecommand \selectlanguage [0]{\@gobble}%
\providecommand \bibinfo  [0]{\@secondoftwo}%
\providecommand \bibfield  [0]{\@secondoftwo}%
\providecommand \translation [1]{[#1]}%
\providecommand \BibitemOpen [0]{}%
\providecommand \bibitemStop [0]{}%
\providecommand \bibitemNoStop [0]{.\EOS\space}%
\providecommand \EOS [0]{\spacefactor3000\relax}%
\providecommand \BibitemShut  [1]{\csname bibitem#1\endcsname}%
\let\auto@bib@innerbib\@empty
\bibitem [{\citenamefont {F{\"{o}}rst}\ \emph {et~al.}(2015)\citenamefont
  {F{\"{o}}rst}, \citenamefont {Mankowsky},\ and\ \citenamefont
  {Cavalleri}}]{Forst2015}%
  \BibitemOpen
  \bibfield  {author} {\bibinfo {author} {\bibfnamefont {M.}~\bibnamefont
  {F{\"{o}}rst}}, \bibinfo {author} {\bibfnamefont {R.}~\bibnamefont
  {Mankowsky}}, \ and\ \bibinfo {author} {\bibfnamefont {A.}~\bibnamefont
  {Cavalleri}},\ }\href@noop {} {\bibfield  {journal} {\bibinfo  {journal}
  {Acc. Chem. Res.}\ }\textbf {\bibinfo {volume} {48}},\ \bibinfo {pages} {380}
  (\bibinfo {year} {2015})}\BibitemShut {NoStop}%
\bibitem [{\citenamefont {Mankowsky}\ \emph {et~al.}(2016)\citenamefont
  {Mankowsky}, \citenamefont {F{\"{o}}rst},\ and\ \citenamefont
  {Cavalleri}}]{Mankowsky2016}%
  \BibitemOpen
  \bibfield  {author} {\bibinfo {author} {\bibfnamefont {R.}~\bibnamefont
  {Mankowsky}}, \bibinfo {author} {\bibfnamefont {M.}~\bibnamefont
  {F{\"{o}}rst}}, \ and\ \bibinfo {author} {\bibfnamefont {A.}~\bibnamefont
  {Cavalleri}},\ }\href@noop {} {\bibfield  {journal} {\bibinfo  {journal}
  {Reports Prog. Phys.}\ }\textbf {\bibinfo {volume} {79}},\ \bibinfo {pages}
  {064503} (\bibinfo {year} {2016})}\BibitemShut {NoStop}%
\bibitem [{\citenamefont {Nicoletti}\ and\ \citenamefont
  {Cavalleri}(2016)}]{Nicoletti2016}%
  \BibitemOpen
  \bibfield  {author} {\bibinfo {author} {\bibfnamefont {D.}~\bibnamefont
  {Nicoletti}}\ and\ \bibinfo {author} {\bibfnamefont {A.}~\bibnamefont
  {Cavalleri}},\ }\href@noop {} {\bibfield  {journal} {\bibinfo  {journal}
  {Adv. Opt. Photonics}\ }\textbf {\bibinfo {volume} {8}},\ \bibinfo {pages}
  {401} (\bibinfo {year} {2016})}\BibitemShut {NoStop}%
\bibitem [{\citenamefont {Stojchevska}\ \emph {et~al.}(2014)\citenamefont
  {Stojchevska}, \citenamefont {Vaskivskyi}, \citenamefont {Mertelj},
  \citenamefont {Kusar}, \citenamefont {Svetin}, \citenamefont {Brazovskii},\
  and\ \citenamefont {Mihailovic}}]{Stojchevska2014}%
  \BibitemOpen
  \bibfield  {author} {\bibinfo {author} {\bibfnamefont {L.}~\bibnamefont
  {Stojchevska}}, \bibinfo {author} {\bibfnamefont {I.}~\bibnamefont
  {Vaskivskyi}}, \bibinfo {author} {\bibfnamefont {T.}~\bibnamefont {Mertelj}},
  \bibinfo {author} {\bibfnamefont {P.}~\bibnamefont {Kusar}}, \bibinfo
  {author} {\bibfnamefont {D.}~\bibnamefont {Svetin}}, \bibinfo {author}
  {\bibfnamefont {S.}~\bibnamefont {Brazovskii}}, \ and\ \bibinfo {author}
  {\bibfnamefont {D.}~\bibnamefont {Mihailovic}},\ }\href@noop {} {\bibfield
  {journal} {\bibinfo  {journal} {Science}\ }\textbf {\bibinfo {volume}
  {344}},\ \bibinfo {pages} {177} (\bibinfo {year} {2014})}\BibitemShut
  {NoStop}%
\bibitem [{\citenamefont {Nova}\ \emph {et~al.}(2016)\citenamefont {Nova},
  \citenamefont {Cartella}, \citenamefont {Cantaluppi}, \citenamefont
  {F{\"{o}}rst}, \citenamefont {Bossini}, \citenamefont {Mikhaylovskiy},
  \citenamefont {Kimel}, \citenamefont {Merlin},\ and\ \citenamefont
  {Cavalleri}}]{Nova2016}%
  \BibitemOpen
  \bibfield  {author} {\bibinfo {author} {\bibfnamefont {T.~F.}\ \bibnamefont
  {Nova}}, \bibinfo {author} {\bibfnamefont {A.}~\bibnamefont {Cartella}},
  \bibinfo {author} {\bibfnamefont {A.}~\bibnamefont {Cantaluppi}}, \bibinfo
  {author} {\bibfnamefont {M.}~\bibnamefont {F{\"{o}}rst}}, \bibinfo {author}
  {\bibfnamefont {D.}~\bibnamefont {Bossini}}, \bibinfo {author} {\bibfnamefont
  {R.~V.}\ \bibnamefont {Mikhaylovskiy}}, \bibinfo {author} {\bibfnamefont
  {A.~V.}\ \bibnamefont {Kimel}}, \bibinfo {author} {\bibfnamefont
  {R.}~\bibnamefont {Merlin}}, \ and\ \bibinfo {author} {\bibfnamefont
  {A.}~\bibnamefont {Cavalleri}},\ }\href@noop {} {\bibfield  {journal}
  {\bibinfo  {journal} {Nat. Phys.}\ }\textbf {\bibinfo {volume} {13}},\
  \bibinfo {pages} {132} (\bibinfo {year} {2016})}\BibitemShut {NoStop}%
\bibitem [{\citenamefont {Rini}\ \emph {et~al.}(2007)\citenamefont {Rini},
  \citenamefont {Tobey}, \citenamefont {Dean}, \citenamefont {Itatani},
  \citenamefont {Tomioka}, \citenamefont {Tokura}, \citenamefont {Schoenlein},\
  and\ \citenamefont {Cavalleri}}]{Rini2007}%
  \BibitemOpen
  \bibfield  {author} {\bibinfo {author} {\bibfnamefont {M.}~\bibnamefont
  {Rini}}, \bibinfo {author} {\bibfnamefont {R.}~\bibnamefont {Tobey}},
  \bibinfo {author} {\bibfnamefont {N.}~\bibnamefont {Dean}}, \bibinfo {author}
  {\bibfnamefont {J.}~\bibnamefont {Itatani}}, \bibinfo {author} {\bibfnamefont
  {Y.}~\bibnamefont {Tomioka}}, \bibinfo {author} {\bibfnamefont
  {Y.}~\bibnamefont {Tokura}}, \bibinfo {author} {\bibfnamefont {R.~W.}\
  \bibnamefont {Schoenlein}}, \ and\ \bibinfo {author} {\bibfnamefont
  {A.}~\bibnamefont {Cavalleri}},\ }\href@noop {} {\bibfield  {journal}
  {\bibinfo  {journal} {Nature} (London)\ }\textbf {\bibinfo {volume} {449}},\ \bibinfo
  {pages} {72} (\bibinfo {year} {2007})}\BibitemShut {NoStop}%
\bibitem [{\citenamefont {Tobey}\ \emph {et~al.}(2008)\citenamefont {Tobey},
  \citenamefont {Prabhakaran}, \citenamefont {Boothroyd},\ and\ \citenamefont
  {Cavalleri}}]{Tobey2008}%
  \BibitemOpen
  \bibfield  {author} {\bibinfo {author} {\bibfnamefont {R.~I.}\ \bibnamefont
  {Tobey}}, \bibinfo {author} {\bibfnamefont {D.}~\bibnamefont {Prabhakaran}},
  \bibinfo {author} {\bibfnamefont {A.~T.}\ \bibnamefont {Boothroyd}}, \ and\
  \bibinfo {author} {\bibfnamefont {A.}~\bibnamefont {Cavalleri}},\ }\href@noop
  {} {\bibfield  {journal} {\bibinfo  {journal} {Phys. Rev. Lett.}\ }\textbf
  {\bibinfo {volume} {101}},\ \bibinfo {pages} {197404} (\bibinfo {year}
  {2008})}\BibitemShut {NoStop}%
\bibitem [{\citenamefont {Fausti}\ \emph {et~al.}(2011)\citenamefont {Fausti},
  \citenamefont {Tobey}, \citenamefont {Dean}, \citenamefont {Kaiser},
  \citenamefont {Dienst}, \citenamefont {Hoffmann}, \citenamefont {Pyon},
  \citenamefont {Takayama}, \citenamefont {Takagi},\ and\ \citenamefont
  {Cavalleri}}]{Fausti2011}%
  \BibitemOpen
  \bibfield  {author} {\bibinfo {author} {\bibfnamefont {D.}~\bibnamefont
  {Fausti}}, \bibinfo {author} {\bibfnamefont {R.~I.}\ \bibnamefont {Tobey}},
  \bibinfo {author} {\bibfnamefont {N.}~\bibnamefont {Dean}}, \bibinfo {author}
  {\bibfnamefont {S.}~\bibnamefont {Kaiser}}, \bibinfo {author} {\bibfnamefont
  {A.}~\bibnamefont {Dienst}}, \bibinfo {author} {\bibfnamefont {M.~C.}\
  \bibnamefont {Hoffmann}}, \bibinfo {author} {\bibfnamefont {S.}~\bibnamefont
  {Pyon}}, \bibinfo {author} {\bibfnamefont {T.}~\bibnamefont {Takayama}},
  \bibinfo {author} {\bibfnamefont {H.}~\bibnamefont {Takagi}}, \ and\ \bibinfo
  {author} {\bibfnamefont {A.}~\bibnamefont {Cavalleri}},\ }\href@noop {}
  {\bibfield  {journal} {\bibinfo  {journal} {Science}\ }\textbf {\bibinfo
  {volume} {331}},\ \bibinfo {pages} {189} (\bibinfo {year}
  {2011})}\BibitemShut {NoStop}%
\bibitem [{\citenamefont {Kaiser}\ \emph {et~al.}(2014)\citenamefont {Kaiser},
  \citenamefont {Hunt}, \citenamefont {Nicoletti}, \citenamefont {Hu},
  \citenamefont {Gierz}, \citenamefont {Liu}, \citenamefont {{Le Tacon}},
  \citenamefont {Loew}, \citenamefont {Haug}, \citenamefont {Keimer},\ and\
  \citenamefont {Cavalleri}}]{Kaiser2014}%
  \BibitemOpen
  \bibfield  {author} {\bibinfo {author} {\bibfnamefont {S.}~\bibnamefont
  {Kaiser}}, \bibinfo {author} {\bibfnamefont {C.~R.}\ \bibnamefont {Hunt}},
  \bibinfo {author} {\bibfnamefont {D.}~\bibnamefont {Nicoletti}}, \bibinfo
  {author} {\bibfnamefont {W.}~\bibnamefont {Hu}}, \bibinfo {author}
  {\bibfnamefont {I.}~\bibnamefont {Gierz}}, \bibinfo {author} {\bibfnamefont
  {H.~Y.}\ \bibnamefont {Liu}}, \bibinfo {author} {\bibfnamefont
  {M.}~\bibnamefont {{Le Tacon}}}, \bibinfo {author} {\bibfnamefont
  {T.}~\bibnamefont {Loew}}, \bibinfo {author} {\bibfnamefont {D.}~\bibnamefont
  {Haug}}, \bibinfo {author} {\bibfnamefont {B.}~\bibnamefont {Keimer}}, \ and\
  \bibinfo {author} {\bibfnamefont {A.}~\bibnamefont {Cavalleri}},\ }\href@noop
  {} {\bibfield  {journal} {\bibinfo  {journal} {Phys. Rev. B}\ }\textbf
  {\bibinfo {volume} {89}},\ \bibinfo {pages} {184516} (\bibinfo {year}
  {2014})}\BibitemShut {NoStop}%
\bibitem [{\citenamefont {Nicoletti}\ \emph {et~al.}(2014)\citenamefont
  {Nicoletti}, \citenamefont {Casandruc}, \citenamefont {Laplace},
  \citenamefont {Khanna}, \citenamefont {Hunt}, \citenamefont {Kaiser},
  \citenamefont {Dhesi}, \citenamefont {Gu}, \citenamefont {Hill},\ and\
  \citenamefont {Cavalleri}}]{Nicoletti2014}%
  \BibitemOpen
  \bibfield  {author} {\bibinfo {author} {\bibfnamefont {D.}~\bibnamefont
  {Nicoletti}}, \bibinfo {author} {\bibfnamefont {E.}~\bibnamefont
  {Casandruc}}, \bibinfo {author} {\bibfnamefont {Y.}~\bibnamefont {Laplace}},
  \bibinfo {author} {\bibfnamefont {V.}~\bibnamefont {Khanna}}, \bibinfo
  {author} {\bibfnamefont {C.~R.}\ \bibnamefont {Hunt}}, \bibinfo {author}
  {\bibfnamefont {S.}~\bibnamefont {Kaiser}}, \bibinfo {author} {\bibfnamefont
  {S.~S.}\ \bibnamefont {Dhesi}}, \bibinfo {author} {\bibfnamefont {G.~D.}\
  \bibnamefont {Gu}}, \bibinfo {author} {\bibfnamefont {J.~P.}\ \bibnamefont
  {Hill}}, \ and\ \bibinfo {author} {\bibfnamefont {A.}~\bibnamefont
  {Cavalleri}},\ }\href@noop {} {\bibfield  {journal} {\bibinfo  {journal}
  {Phys. Rev. B}\ }\textbf {\bibinfo {volume} {90}},\ \bibinfo {pages} {100503}
  (\bibinfo {year} {2014})}\BibitemShut {NoStop}%
\bibitem [{\citenamefont {F{\"{o}}rst}\ \emph
  {et~al.}(2014{\natexlab{a}})\citenamefont {F{\"{o}}rst}, \citenamefont
  {Frano}, \citenamefont {Kaiser}, \citenamefont {Mankowsky}, \citenamefont
  {Hunt}, \citenamefont {Turner}, \citenamefont {Dakovski}, \citenamefont
  {Minitti}, \citenamefont {Robinson}, \citenamefont {Loew}, \citenamefont {{Le
  Tacon}}, \citenamefont {Keimer}, \citenamefont {Hill}, \citenamefont
  {Cavalleri},\ and\ \citenamefont {Dhesi}}]{Forst2014}%
  \BibitemOpen
  \bibfield  {author} {\bibinfo {author} {\bibfnamefont {M.}~\bibnamefont
  {F{\"{o}}rst}}, \bibinfo {author} {\bibfnamefont {A.}~\bibnamefont {Frano}},
  \bibinfo {author} {\bibfnamefont {S.}~\bibnamefont {Kaiser}}, \bibinfo
  {author} {\bibfnamefont {R.}~\bibnamefont {Mankowsky}}, \bibinfo {author}
  {\bibfnamefont {C.~R.}\ \bibnamefont {Hunt}}, \bibinfo {author}
  {\bibfnamefont {J.~J.}\ \bibnamefont {Turner}}, \bibinfo {author}
  {\bibfnamefont {G.~L.}\ \bibnamefont {Dakovski}}, \bibinfo {author}
  {\bibfnamefont {M.~P.}\ \bibnamefont {Minitti}}, \bibinfo {author}
  {\bibfnamefont {J.}~\bibnamefont {Robinson}}, \bibinfo {author}
  {\bibfnamefont {T.}~\bibnamefont {Loew}}, \bibinfo {author} {\bibfnamefont
  {M.}~\bibnamefont {{Le Tacon}}}, \bibinfo {author} {\bibfnamefont
  {B.}~\bibnamefont {Keimer}}, \bibinfo {author} {\bibfnamefont {J.~P.}\
  \bibnamefont {Hill}}, \bibinfo {author} {\bibfnamefont {A.}~\bibnamefont
  {Cavalleri}}, \ and\ \bibinfo {author} {\bibfnamefont {S.~S.}\ \bibnamefont
  {Dhesi}},\ }\href@noop {} {\bibfield  {journal} {\bibinfo  {journal} {Phys.
  Rev. B}\ }\textbf {\bibinfo {volume} {90}},\ \bibinfo {pages} {184514}
  (\bibinfo {year} {2014}{\natexlab{a}})}\BibitemShut {NoStop}%
\bibitem [{\citenamefont {F{\"{o}}rst}\ \emph
  {et~al.}(2014{\natexlab{b}})\citenamefont {F{\"{o}}rst}, \citenamefont
  {Tobey}, \citenamefont {Bromberger}, \citenamefont {Wilkins}, \citenamefont
  {Khanna}, \citenamefont {Caviglia}, \citenamefont {Chuang}, \citenamefont
  {Lee}, \citenamefont {Schlotter}, \citenamefont {Turner}, \citenamefont
  {Minitti}, \citenamefont {Krupin}, \citenamefont {Xu}, \citenamefont {Wen},
  \citenamefont {Gu}, \citenamefont {Dhesi}, \citenamefont {Cavalleri},\ and\
  \citenamefont {Hill}}]{Forst2014a}%
  \BibitemOpen
  \bibfield  {author} {\bibinfo {author} {\bibfnamefont {M.}~\bibnamefont
  {F{\"{o}}rst}}, \bibinfo {author} {\bibfnamefont {R.~I.}\ \bibnamefont
  {Tobey}}, \bibinfo {author} {\bibfnamefont {H.}~\bibnamefont {Bromberger}},
  \bibinfo {author} {\bibfnamefont {S.~B.}\ \bibnamefont {Wilkins}}, \bibinfo
  {author} {\bibfnamefont {V.}~\bibnamefont {Khanna}}, \bibinfo {author}
  {\bibfnamefont {A.~D.}\ \bibnamefont {Caviglia}}, \bibinfo {author}
  {\bibfnamefont {Y.~D.}\ \bibnamefont {Chuang}}, \bibinfo {author}
  {\bibfnamefont {W.~S.}\ \bibnamefont {Lee}}, \bibinfo {author} {\bibfnamefont
  {W.~F.}\ \bibnamefont {Schlotter}}, \bibinfo {author} {\bibfnamefont {J.~J.}\
  \bibnamefont {Turner}}, \bibinfo {author} {\bibfnamefont {M.~P.}\
  \bibnamefont {Minitti}}, \bibinfo {author} {\bibfnamefont {O.}~\bibnamefont
  {Krupin}}, \bibinfo {author} {\bibfnamefont {Z.~J.}\ \bibnamefont {Xu}},
  \bibinfo {author} {\bibfnamefont {J.~S.}\ \bibnamefont {Wen}}, \bibinfo
  {author} {\bibfnamefont {G.~D.}\ \bibnamefont {Gu}}, \bibinfo {author}
  {\bibfnamefont {S.~S.}\ \bibnamefont {Dhesi}}, \bibinfo {author}
  {\bibfnamefont {A.}~\bibnamefont {Cavalleri}}, \ and\ \bibinfo {author}
  {\bibfnamefont {J.~P.}\ \bibnamefont {Hill}},\ }\href@noop {} {\bibfield
  {journal} {\bibinfo  {journal} {Phys. Rev. Lett.}\ }\textbf {\bibinfo
  {volume} {112}},\ \bibinfo {pages} {157002} (\bibinfo {year}
  {2014}{\natexlab{b}})}\BibitemShut {NoStop}%
\bibitem [{\citenamefont {Hu}\ \emph {et~al.}(2014)\citenamefont {Hu},
  \citenamefont {Kaiser}, \citenamefont {Nicoletti}, \citenamefont {Hunt},
  \citenamefont {Gierz}, \citenamefont {Hoffmann}, \citenamefont {{Le Tacon}},
  \citenamefont {Loew}, \citenamefont {Keimer},\ and\ \citenamefont
  {Cavalleri}}]{Hu2014}%
  \BibitemOpen
  \bibfield  {author} {\bibinfo {author} {\bibfnamefont {W.}~\bibnamefont
  {Hu}}, \bibinfo {author} {\bibfnamefont {S.}~\bibnamefont {Kaiser}}, \bibinfo
  {author} {\bibfnamefont {D.}~\bibnamefont {Nicoletti}}, \bibinfo {author}
  {\bibfnamefont {C.~R.}\ \bibnamefont {Hunt}}, \bibinfo {author}
  {\bibfnamefont {I.}~\bibnamefont {Gierz}}, \bibinfo {author} {\bibfnamefont
  {M.~C.}\ \bibnamefont {Hoffmann}}, \bibinfo {author} {\bibfnamefont
  {M.}~\bibnamefont {{Le Tacon}}}, \bibinfo {author} {\bibfnamefont
  {T.}~\bibnamefont {Loew}}, \bibinfo {author} {\bibfnamefont {B.}~\bibnamefont
  {Keimer}}, \ and\ \bibinfo {author} {\bibfnamefont {A.}~\bibnamefont
  {Cavalleri}},\ }\href@noop {} {\bibfield  {journal} {\bibinfo  {journal}
  {Nat. Mater.}\ }\textbf {\bibinfo {volume} {13}},\ \bibinfo {pages} {705}
  (\bibinfo {year} {2014})}\BibitemShut {NoStop}%
\bibitem [{\citenamefont {Mankowsky}\ \emph {et~al.}(2014)\citenamefont
  {Mankowsky}, \citenamefont {Subedi}, \citenamefont {F{\"{o}}rst},
  \citenamefont {Mariager}, \citenamefont {Chollet}, \citenamefont {Lemke},
  \citenamefont {Robinson}, \citenamefont {Glownia}, \citenamefont {Minitti},
  \citenamefont {Frano}, \citenamefont {Fechner}, \citenamefont {Spaldin},
  \citenamefont {Loew}, \citenamefont {Keimer}, \citenamefont {Georges},\ and\
  \citenamefont {Cavalleri}}]{Mankowsky2014}%
  \BibitemOpen
  \bibfield  {author} {\bibinfo {author} {\bibfnamefont {R.}~\bibnamefont
  {Mankowsky}}, \bibinfo {author} {\bibfnamefont {A.}~\bibnamefont {Subedi}},
  \bibinfo {author} {\bibfnamefont {M.}~\bibnamefont {F{\"{o}}rst}}, \bibinfo
  {author} {\bibfnamefont {S.~O.}\ \bibnamefont {Mariager}}, \bibinfo {author}
  {\bibfnamefont {M.}~\bibnamefont {Chollet}}, \bibinfo {author} {\bibfnamefont
  {H.~T.}\ \bibnamefont {Lemke}}, \bibinfo {author} {\bibfnamefont {J.~S.}\
  \bibnamefont {Robinson}}, \bibinfo {author} {\bibfnamefont {J.~M.}\
  \bibnamefont {Glownia}}, \bibinfo {author} {\bibfnamefont {M.~P.}\
  \bibnamefont {Minitti}}, \bibinfo {author} {\bibfnamefont {A.}~\bibnamefont
  {Frano}}, \bibinfo {author} {\bibfnamefont {M.}~\bibnamefont {Fechner}},
  \bibinfo {author} {\bibfnamefont {N.~A.}\ \bibnamefont {Spaldin}}, \bibinfo
  {author} {\bibfnamefont {T.}~\bibnamefont {Loew}}, \bibinfo {author}
  {\bibfnamefont {B.}~\bibnamefont {Keimer}}, \bibinfo {author} {\bibfnamefont
  {A.}~\bibnamefont {Georges}}, \ and\ \bibinfo {author} {\bibfnamefont
  {A.}~\bibnamefont {Cavalleri}},\ }\href@noop {} {\bibfield  {journal}
  {\bibinfo  {journal} {Nature (London)}\ }\textbf {\bibinfo {volume} {516}},\ \bibinfo
  {pages} {71} (\bibinfo {year} {2014})}\BibitemShut {NoStop}%
\bibitem [{\citenamefont {Hunt}\ \emph {et~al.}(2015)\citenamefont {Hunt},
  \citenamefont {Nicoletti}, \citenamefont {Kaiser}, \citenamefont {Takayama},
  \citenamefont {Takagi},\ and\ \citenamefont {Cavalleri}}]{Hunt2015}%
  \BibitemOpen
  \bibfield  {author} {\bibinfo {author} {\bibfnamefont {C.~R.}\ \bibnamefont
  {Hunt}}, \bibinfo {author} {\bibfnamefont {D.}~\bibnamefont {Nicoletti}},
  \bibinfo {author} {\bibfnamefont {S.}~\bibnamefont {Kaiser}}, \bibinfo
  {author} {\bibfnamefont {T.}~\bibnamefont {Takayama}}, \bibinfo {author}
  {\bibfnamefont {H.}~\bibnamefont {Takagi}}, \ and\ \bibinfo {author}
  {\bibfnamefont {A.}~\bibnamefont {Cavalleri}},\ }\href@noop {} {\bibfield
  {journal} {\bibinfo  {journal} {Phys. Rev. B}\ }\textbf {\bibinfo {volume}
  {91}},\ \bibinfo {pages} {020505} (\bibinfo {year} {2015})}\BibitemShut
  {NoStop}%
\bibitem [{\citenamefont {Casandruc}\ \emph {et~al.}(2015)\citenamefont
  {Casandruc}, \citenamefont {Nicoletti}, \citenamefont {Rajasekaran},
  \citenamefont {Laplace}, \citenamefont {Khanna}, \citenamefont {Gu},
  \citenamefont {Hill},\ and\ \citenamefont {Cavalleri}}]{Casandruc2015}%
  \BibitemOpen
  \bibfield  {author} {\bibinfo {author} {\bibfnamefont {E.}~\bibnamefont
  {Casandruc}}, \bibinfo {author} {\bibfnamefont {D.}~\bibnamefont
  {Nicoletti}}, \bibinfo {author} {\bibfnamefont {S.}~\bibnamefont
  {Rajasekaran}}, \bibinfo {author} {\bibfnamefont {Y.}~\bibnamefont
  {Laplace}}, \bibinfo {author} {\bibfnamefont {V.}~\bibnamefont {Khanna}},
  \bibinfo {author} {\bibfnamefont {G.~D.}\ \bibnamefont {Gu}}, \bibinfo
  {author} {\bibfnamefont {J.~P.}\ \bibnamefont {Hill}}, \ and\ \bibinfo
  {author} {\bibfnamefont {A.}~\bibnamefont {Cavalleri}},\ }\href@noop {}
  {\bibfield  {journal} {\bibinfo  {journal} {Phys. Rev. B}\ }\textbf {\bibinfo
  {volume} {91}},\ \bibinfo {pages} {174502} (\bibinfo {year}
  {2015})}\BibitemShut {NoStop}%
\bibitem [{\citenamefont {Mankowsky}\ \emph {et~al.}(2015)\citenamefont
  {Mankowsky}, \citenamefont {F{\"{o}}rst}, \citenamefont {Loew}, \citenamefont
  {Porras}, \citenamefont {Keimer},\ and\ \citenamefont
  {Cavalleri}}]{Mankowsky2015}%
  \BibitemOpen
  \bibfield  {author} {\bibinfo {author} {\bibfnamefont {R.}~\bibnamefont
  {Mankowsky}}, \bibinfo {author} {\bibfnamefont {M.}~\bibnamefont
  {F{\"{o}}rst}}, \bibinfo {author} {\bibfnamefont {T.}~\bibnamefont {Loew}},
  \bibinfo {author} {\bibfnamefont {J.}~\bibnamefont {Porras}}, \bibinfo
  {author} {\bibfnamefont {B.}~\bibnamefont {Keimer}}, \ and\ \bibinfo {author}
  {\bibfnamefont {A.}~\bibnamefont {Cavalleri}},\ }\href@noop {} {\bibfield
  {journal} {\bibinfo  {journal} {Phys. Rev. B}\ }\textbf {\bibinfo {volume}
  {91}},\ \bibinfo {pages} {094308} (\bibinfo {year} {2015})}\BibitemShut
  {NoStop}%
\bibitem [{\citenamefont {Khanna}\ \emph {et~al.}(2016)\citenamefont {Khanna},
  \citenamefont {Mankowsky}, \citenamefont {Petrich}, \citenamefont
  {Bromberger}, \citenamefont {Cavill}, \citenamefont {M{\"{o}}hr-Vorobeva},
  \citenamefont {Nicoletti}, \citenamefont {Laplace}, \citenamefont {Gu},
  \citenamefont {Hill}, \citenamefont {F{\"{o}}rst}, \citenamefont
  {Cavalleri},\ and\ \citenamefont {Dhesi}}]{Khanna2016}%
  \BibitemOpen
  \bibfield  {author} {\bibinfo {author} {\bibfnamefont {V.}~\bibnamefont
  {Khanna}}, \bibinfo {author} {\bibfnamefont {R.}~\bibnamefont {Mankowsky}},
  \bibinfo {author} {\bibfnamefont {M.}~\bibnamefont {Petrich}}, \bibinfo
  {author} {\bibfnamefont {H.}~\bibnamefont {Bromberger}}, \bibinfo {author}
  {\bibfnamefont {S.~A.}\ \bibnamefont {Cavill}}, \bibinfo {author}
  {\bibfnamefont {E.}~\bibnamefont {M{\"{o}}hr-Vorobeva}}, \bibinfo {author}
  {\bibfnamefont {D.}~\bibnamefont {Nicoletti}}, \bibinfo {author}
  {\bibfnamefont {Y.}~\bibnamefont {Laplace}}, \bibinfo {author} {\bibfnamefont
  {G.~D.}\ \bibnamefont {Gu}}, \bibinfo {author} {\bibfnamefont {J.~P.}\
  \bibnamefont {Hill}}, \bibinfo {author} {\bibfnamefont {M.}~\bibnamefont
  {F{\"{o}}rst}}, \bibinfo {author} {\bibfnamefont {A.}~\bibnamefont
  {Cavalleri}}, \ and\ \bibinfo {author} {\bibfnamefont {S.~S.}\ \bibnamefont
  {Dhesi}},\ }\href@noop {} {\bibfield  {journal} {\bibinfo  {journal} {Phys.
  Rev. B}\ }\textbf {\bibinfo {volume} {93}},\ \bibinfo {pages} {224522}
  (\bibinfo {year} {2016})}\BibitemShut {NoStop}%
\bibitem [{\citenamefont {Hunt}\ \emph {et~al.}(2016)\citenamefont {Hunt},
  \citenamefont {Nicoletti}, \citenamefont {Kaiser}, \citenamefont
  {Pr{\"{o}}pper}, \citenamefont {Loew}, \citenamefont {Porras}, \citenamefont
  {Keimer},\ and\ \citenamefont {Cavalleri}}]{Hunt2016}%
  \BibitemOpen
  \bibfield  {author} {\bibinfo {author} {\bibfnamefont {C.~R.}\ \bibnamefont
  {Hunt}}, \bibinfo {author} {\bibfnamefont {D.}~\bibnamefont {Nicoletti}},
  \bibinfo {author} {\bibfnamefont {S.}~\bibnamefont {Kaiser}}, \bibinfo
  {author} {\bibfnamefont {D.}~\bibnamefont {Pr{\"{o}}pper}}, \bibinfo {author}
  {\bibfnamefont {T.}~\bibnamefont {Loew}}, \bibinfo {author} {\bibfnamefont
  {J.}~\bibnamefont {Porras}}, \bibinfo {author} {\bibfnamefont
  {B.}~\bibnamefont {Keimer}}, \ and\ \bibinfo {author} {\bibfnamefont
  {A.}~\bibnamefont {Cavalleri}},\ }\href@noop {} {\bibfield  {journal}
  {\bibinfo  {journal} {Phys. Rev. B}\ }\textbf {\bibinfo {volume} {94}},\
  \bibinfo {pages} {224303} (\bibinfo {year} {2016})}\BibitemShut {NoStop}%
\bibitem [{\citenamefont {Mankowsky}\ \emph {et~al.}(2017)\citenamefont
  {Mankowsky}, \citenamefont {Fechner}, \citenamefont {F{\"{o}}rst},
  \citenamefont {von Hoegen}, \citenamefont {Porras}, \citenamefont {Loew},
  \citenamefont {Dakovski}, \citenamefont {Seaberg}, \citenamefont
  {M{\"{o}}ller}, \citenamefont {Coslovich}, \citenamefont {Keimer},
  \citenamefont {Dhesi},\ and\ \citenamefont {Cavalleri}}]{Mankowsky2017}%
  \BibitemOpen
  \bibfield  {author} {\bibinfo {author} {\bibfnamefont {R.}~\bibnamefont
  {Mankowsky}}, \bibinfo {author} {\bibfnamefont {M.}~\bibnamefont {Fechner}},
  \bibinfo {author} {\bibfnamefont {M.}~\bibnamefont {F{\"{o}}rst}}, \bibinfo
  {author} {\bibfnamefont {A.}~\bibnamefont {von Hoegen}}, \bibinfo {author}
  {\bibfnamefont {J.}~\bibnamefont {Porras}}, \bibinfo {author} {\bibfnamefont
  {T.}~\bibnamefont {Loew}}, \bibinfo {author} {\bibfnamefont {G.~L.}\
  \bibnamefont {Dakovski}}, \bibinfo {author} {\bibfnamefont {M.}~\bibnamefont
  {Seaberg}}, \bibinfo {author} {\bibfnamefont {S.}~\bibnamefont
  {M{\"{o}}ller}}, \bibinfo {author} {\bibfnamefont {G.}~\bibnamefont
  {Coslovich}}, \bibinfo {author} {\bibfnamefont {B.}~\bibnamefont {Keimer}},
  \bibinfo {author} {\bibfnamefont {S.~S.}\ \bibnamefont {Dhesi}}, \ and\
  \bibinfo {author} {\bibfnamefont {A.}~\bibnamefont {Cavalleri}},\ }\href@noop
  {} {\bibfield  {journal} {\bibinfo  {journal} {Struct. Dyn.}\ }\textbf
  {\bibinfo {volume} {4}},\ \bibinfo {pages} {044007} (\bibinfo {year}
  {2017})}\BibitemShut {NoStop}%
\bibitem [{\citenamefont {Hu}\ \emph {et~al.}(2017)\citenamefont {Hu},
  \citenamefont {Nicoletti}, \citenamefont {Boris}, \citenamefont {Keimer},\
  and\ \citenamefont {Cavalleri}}]{Hu2017}%
  \BibitemOpen
  \bibfield  {author} {\bibinfo {author} {\bibfnamefont {W.}~\bibnamefont
  {Hu}}, \bibinfo {author} {\bibfnamefont {D.}~\bibnamefont {Nicoletti}},
  \bibinfo {author} {\bibfnamefont {A.~V.}\ \bibnamefont {Boris}}, \bibinfo
  {author} {\bibfnamefont {B.}~\bibnamefont {Keimer}}, \ and\ \bibinfo {author}
  {\bibfnamefont {A.}~\bibnamefont {Cavalleri}},\ }\href@noop {} {\bibfield
  {journal} {\bibinfo  {journal} {Phys. Rev. B}\ }\textbf {\bibinfo {volume}
  {95}},\ \bibinfo {pages} {104508} (\bibinfo {year} {2017})}\BibitemShut
  {NoStop}%
\bibitem [{\citenamefont {Denny}\ \emph {et~al.}(2015)\citenamefont {Denny},
  \citenamefont {Clark}, \citenamefont {Laplace}, \citenamefont {Cavalleri},\
  and\ \citenamefont {Jaksch}}]{Denny2015}%
  \BibitemOpen
  \bibfield  {author} {\bibinfo {author} {\bibfnamefont {S.~J.}\ \bibnamefont
  {Denny}}, \bibinfo {author} {\bibfnamefont {S.~R.}\ \bibnamefont {Clark}},
  \bibinfo {author} {\bibfnamefont {Y.}~\bibnamefont {Laplace}}, \bibinfo
  {author} {\bibfnamefont {A.}~\bibnamefont {Cavalleri}}, \ and\ \bibinfo
  {author} {\bibfnamefont {D.}~\bibnamefont {Jaksch}},\ }\href@noop {}
  {\bibfield  {journal} {\bibinfo  {journal} {Phys. Rev. Lett.}\ }\textbf
  {\bibinfo {volume} {114}},\ \bibinfo {pages} {137001} (\bibinfo {year}
  {2015})}\BibitemShut {NoStop}%
\bibitem [{\citenamefont {Raines}\ \emph {et~al.}(2015)\citenamefont {Raines},
  \citenamefont {Stanev},\ and\ \citenamefont {Galitski}}]{Raines2015}%
  \BibitemOpen
  \bibfield  {author} {\bibinfo {author} {\bibfnamefont {Z.~M.}\ \bibnamefont
  {Raines}}, \bibinfo {author} {\bibfnamefont {V.}~\bibnamefont {Stanev}}, \
  and\ \bibinfo {author} {\bibfnamefont {V.~M.}\ \bibnamefont {Galitski}},\
  }\href@noop {} {\bibfield  {journal} {\bibinfo  {journal} {Phys. Rev. B}\
  }\textbf {\bibinfo {volume} {91}},\ \bibinfo {pages} {184506} (\bibinfo
  {year} {2015})}\BibitemShut {NoStop}%
\bibitem [{\citenamefont {H{\"{o}}ppner}\ \emph {et~al.}(2015)\citenamefont
  {H{\"{o}}ppner}, \citenamefont {Zhu}, \citenamefont {Rexin}, \citenamefont
  {Cavalleri},\ and\ \citenamefont {Mathey}}]{Hoppner2015}%
  \BibitemOpen
  \bibfield  {author} {\bibinfo {author} {\bibfnamefont {R.}~\bibnamefont
  {H{\"{o}}ppner}}, \bibinfo {author} {\bibfnamefont {B.}~\bibnamefont {Zhu}},
  \bibinfo {author} {\bibfnamefont {T.}~\bibnamefont {Rexin}}, \bibinfo
  {author} {\bibfnamefont {A.}~\bibnamefont {Cavalleri}}, \ and\ \bibinfo
  {author} {\bibfnamefont {L.}~\bibnamefont {Mathey}},\ }\href@noop {}
  {\bibfield  {journal} {\bibinfo  {journal} {Phys. Rev. B}\ }\textbf {\bibinfo
  {volume} {91}},\ \bibinfo {pages} {104507} (\bibinfo {year}
  {2015})}\BibitemShut {NoStop}%
\bibitem [{\citenamefont {Patel}\ and\ \citenamefont
  {Eberlein}(2016)}]{Patel2016}%
  \BibitemOpen
  \bibfield  {author} {\bibinfo {author} {\bibfnamefont {A.~A.}\ \bibnamefont
  {Patel}}\ and\ \bibinfo {author} {\bibfnamefont {A.}~\bibnamefont
  {Eberlein}},\ }\href@noop {} {\bibfield  {journal} {\bibinfo  {journal}
  {Phys. Rev. B}\ }\textbf {\bibinfo {volume} {93}},\ \bibinfo {pages} {195139}
  (\bibinfo {year} {2016})}\BibitemShut {NoStop}%
\bibitem [{\citenamefont {Okamoto}\ \emph {et~al.}(2016)\citenamefont
  {Okamoto}, \citenamefont {Cavalleri},\ and\ \citenamefont
  {Mathey}}]{Okamoto2016}%
  \BibitemOpen
  \bibfield  {author} {\bibinfo {author} {\bibfnamefont {J.~I.}\ \bibnamefont
  {Okamoto}}, \bibinfo {author} {\bibfnamefont {A.}~\bibnamefont {Cavalleri}},
  \ and\ \bibinfo {author} {\bibfnamefont {L.}~\bibnamefont {Mathey}},\
  }\href@noop {} {\bibfield  {journal} {\bibinfo  {journal} {Phys. Rev. Lett.}\
  }\textbf {\bibinfo {volume} {117}},\ \bibinfo {pages} {227001} (\bibinfo
  {year} {2016})}\BibitemShut {NoStop}%
\bibitem [{\citenamefont {Sentef}\ \emph {et~al.}(2017)\citenamefont {Sentef},
  \citenamefont {Tokuno}, \citenamefont {Georges},\ and\ \citenamefont
  {Kollath}}]{Sentef2017}%
  \BibitemOpen
  \bibfield  {author} {\bibinfo {author} {\bibfnamefont {M.~A.}\ \bibnamefont
  {Sentef}}, \bibinfo {author} {\bibfnamefont {A.}~\bibnamefont {Tokuno}},
  \bibinfo {author} {\bibfnamefont {A.}~\bibnamefont {Georges}}, \ and\
  \bibinfo {author} {\bibfnamefont {C.}~\bibnamefont {Kollath}},\ }\href@noop
  {} {\bibfield  {journal} {\bibinfo  {journal} {Phys. Rev. Lett.}\ }\textbf
  {\bibinfo {volume} {118}},\ \bibinfo {pages} {087002} (\bibinfo {year}
  {2017})}\BibitemShut {NoStop}%
\bibitem [{\citenamefont {van~der Marel}\ and\ \citenamefont
  {Tsvetkov}(1996)}]{VanderMarel1996}%
  \BibitemOpen
  \bibfield  {author} {\bibinfo {author} {\bibfnamefont {D.}~\bibnamefont
  {van~der Marel}}\ and\ \bibinfo {author} {\bibfnamefont {a.}~\bibnamefont
  {Tsvetkov}},\ }\href@noop {} {\bibfield  {journal} {\bibinfo  {journal}
  {Czechoslov. J. Phys.}\ }\textbf {\bibinfo {volume} {46}},\ \bibinfo {pages}
  {3165} (\bibinfo {year} {1996})}\BibitemShut {NoStop}%
\bibitem [{\citenamefont {Koyama}\ and\ \citenamefont
  {Tachiki}(1996)}]{Koyama1996}%
  \BibitemOpen
  \bibfield  {author} {\bibinfo {author} {\bibfnamefont {T.}~\bibnamefont
  {Koyama}}\ and\ \bibinfo {author} {\bibfnamefont {M.}~\bibnamefont
  {Tachiki}},\ }\href@noop {} {\bibfield  {journal} {\bibinfo  {journal} {Phys.
  Rev. B}\ }\textbf {\bibinfo {volume} {54}},\ \bibinfo {pages} {16183}
  (\bibinfo {year} {1996})}\BibitemShut {NoStop}%
\bibitem [{\citenamefont {Matsumoto}\ \emph {et~al.}(1999)\citenamefont
  {Matsumoto}, \citenamefont {Sakamoto}, \citenamefont {Wajima}, \citenamefont
  {Koyama},\ and\ \citenamefont {Machida}}]{Matsumoto1999}%
  \BibitemOpen
  \bibfield  {author} {\bibinfo {author} {\bibfnamefont {H.}~\bibnamefont
  {Matsumoto}}, \bibinfo {author} {\bibfnamefont {S.}~\bibnamefont {Sakamoto}},
  \bibinfo {author} {\bibfnamefont {F.}~\bibnamefont {Wajima}}, \bibinfo
  {author} {\bibfnamefont {T.}~\bibnamefont {Koyama}}, \ and\ \bibinfo {author}
  {\bibfnamefont {M.}~\bibnamefont {Machida}},\ }\href@noop {} {\bibfield
  {journal} {\bibinfo  {journal} {Phys. Rev. B}\ }\textbf {\bibinfo {volume}
  {60}},\ \bibinfo {pages} {3666} (\bibinfo {year} {1999})}\BibitemShut
  {NoStop}%
\bibitem [{\citenamefont {Koyama}(1999)}]{Koyama1999}%
  \BibitemOpen
  \bibfield  {author} {\bibinfo {author} {\bibfnamefont {T.}~\bibnamefont
  {Koyama}},\ }\href@noop {} {\bibfield  {journal} {\bibinfo  {journal} {J.
  Phys. Soc. Japan}\ }\textbf {\bibinfo {volume} {68}},\ \bibinfo {pages}
  {3062} (\bibinfo {year} {1999})}\BibitemShut {NoStop}%
\bibitem [{\citenamefont {Machida}\ \emph {et~al.}(1999)\citenamefont
  {Machida}, \citenamefont {Koyama},\ and\ \citenamefont
  {Tachiki}}]{Machida1999}%
  \BibitemOpen
  \bibfield  {author} {\bibinfo {author} {\bibfnamefont {M.}~\bibnamefont
  {Machida}}, \bibinfo {author} {\bibfnamefont {T.}~\bibnamefont {Koyama}}, \
  and\ \bibinfo {author} {\bibfnamefont {M.}~\bibnamefont {Tachiki}},\
  }\href@noop {} {\bibfield  {journal} {\bibinfo  {journal} {Phys. Rev. Lett.}\
  }\textbf {\bibinfo {volume} {83}},\ \bibinfo {pages} {4618} (\bibinfo {year}
  {1999})}\BibitemShut {NoStop}%
\bibitem [{\citenamefont {Koyama}(2000)}]{Koyama2000}%
  \BibitemOpen
  \bibfield  {author} {\bibinfo {author} {\bibfnamefont {T.}~\bibnamefont
  {Koyama}},\ }\href@noop {} {\bibfield  {journal} {\bibinfo  {journal} {J.
  Phys. Soc. Japan}\ }\textbf {\bibinfo {volume} {69}},\ \bibinfo {pages}
  {3689} (\bibinfo {year} {2000})}\BibitemShut {NoStop}%
\bibitem [{\citenamefont {van~der Marel}\ and\ \citenamefont
  {Tsvetkov}(2001)}]{VanderMarel2001}%
  \BibitemOpen
  \bibfield  {author} {\bibinfo {author} {\bibfnamefont {D.}~\bibnamefont
  {van~der Marel}}\ and\ \bibinfo {author} {\bibfnamefont {A.~A.}\ \bibnamefont
  {Tsvetkov}},\ }\href@noop {} {\bibfield  {journal} {\bibinfo  {journal}
  {Phys. Rev. B}\ }\textbf {\bibinfo {volume} {64}},\ \bibinfo {pages} {024530}
  (\bibinfo {year} {2001})}\BibitemShut {NoStop}%
\bibitem [{\citenamefont {Koyama}(2001)}]{Koyama2001}%
  \BibitemOpen
  \bibfield  {author} {\bibinfo {author} {\bibfnamefont {T.}~\bibnamefont
  {Koyama}},\ }\href@noop {} {\bibfield  {journal} {\bibinfo  {journal} {J.
  Phys. Soc. Japan}\ }\textbf {\bibinfo {volume} {70}},\ \bibinfo {pages}
  {2114} (\bibinfo {year} {2001})}\BibitemShut {NoStop}%
\bibitem [{\citenamefont {Koyama}(2002)}]{Koyama2002}%
  \BibitemOpen
  \bibfield  {author} {\bibinfo {author} {\bibfnamefont {T.}~\bibnamefont
  {Koyama}},\ }\href@noop {} {\bibfield  {journal} {\bibinfo  {journal} {J.
  Phys. Soc. Japan}\ }\textbf {\bibinfo {volume} {71}},\ \bibinfo {pages}
  {2986} (\bibinfo {year} {2002})}\BibitemShut {NoStop}%
\bibitem [{\citenamefont {Machida}\ and\ \citenamefont
  {Koyama}(2004)}]{Machida2004}%
  \BibitemOpen
  \bibfield  {author} {\bibinfo {author} {\bibfnamefont {M.}~\bibnamefont
  {Machida}}\ and\ \bibinfo {author} {\bibfnamefont {T.}~\bibnamefont
  {Koyama}},\ }\href@noop {} {\bibfield  {journal} {\bibinfo  {journal} {Phys.
  Rev. B}\ }\textbf {\bibinfo {volume} {70}},\ \bibinfo {pages} {024523}
  (\bibinfo {year} {2004})}\BibitemShut {NoStop}%
\bibitem [{\citenamefont {Machida}\ and\ \citenamefont
  {Sakai}(2004)}]{Machida2004a}%
  \BibitemOpen
  \bibfield  {author} {\bibinfo {author} {\bibfnamefont {M.}~\bibnamefont
  {Machida}}\ and\ \bibinfo {author} {\bibfnamefont {S.}~\bibnamefont
  {Sakai}},\ }\href@noop {} {\bibfield  {journal} {\bibinfo  {journal} {Phys.
  Rev. B}\ }\textbf {\bibinfo {volume} {70}},\ \bibinfo {pages} {144520}
  (\bibinfo {year} {2004})}\BibitemShut {NoStop}%
\bibitem [{\citenamefont {Shukrinov}\ and\ \citenamefont
  {Mahfouzi}(2007)}]{Shukrinov2007}%
  \BibitemOpen
  \bibfield  {author} {\bibinfo {author} {\bibfnamefont {Y.~M.}\ \bibnamefont
  {Shukrinov}}\ and\ \bibinfo {author} {\bibfnamefont {F.}~\bibnamefont
  {Mahfouzi}},\ }\href@noop {} {\bibfield  {journal} {\bibinfo  {journal}
  {Phys. Rev. Lett.}\ }\textbf {\bibinfo {volume} {98}},\ \bibinfo {pages}
  {157001} (\bibinfo {year} {2007})}\BibitemShut {NoStop}%
\bibitem [{\citenamefont {Shukrinov}\ \emph {et~al.}(2009)\citenamefont
  {Shukrinov}, \citenamefont {Hamdipour},\ and\ \citenamefont
  {Kolahchi}}]{Shukrinov2009}%
  \BibitemOpen
  \bibfield  {author} {\bibinfo {author} {\bibfnamefont {Y.~M.}\ \bibnamefont
  {Shukrinov}}, \bibinfo {author} {\bibfnamefont {M.}~\bibnamefont
  {Hamdipour}}, \ and\ \bibinfo {author} {\bibfnamefont {M.~R.}\ \bibnamefont
  {Kolahchi}},\ }\href@noop {} {\bibfield  {journal} {\bibinfo  {journal}
  {Phys. Rev. B}\ }\textbf {\bibinfo {volume} {80}},\ \bibinfo {pages} {014512}
  (\bibinfo {year} {2009})}\BibitemShut {NoStop}%
\bibitem [{\citenamefont {Jung}(1993)}]{Jung1993}%
  \BibitemOpen
  \bibfield  {author} {\bibinfo {author} {\bibfnamefont {P.}~\bibnamefont
  {Jung}},\ }\href@noop {} {\bibfield  {journal} {\bibinfo  {journal} {Phys.
  Rep.}\ }\textbf {\bibinfo {volume} {234}},\ \bibinfo {pages} {175} (\bibinfo
  {year} {1993})}\BibitemShut {NoStop}%
\bibitem [{\citenamefont {Zerbe}\ \emph {et~al.}(1994)\citenamefont {Zerbe},
  \citenamefont {Jung},\ and\ \citenamefont {H{\"{a}}nggi}}]{Zerbe1994}%
  \BibitemOpen
  \bibfield  {author} {\bibinfo {author} {\bibfnamefont {C.}~\bibnamefont
  {Zerbe}}, \bibinfo {author} {\bibfnamefont {P.}~\bibnamefont {Jung}}, \ and\
  \bibinfo {author} {\bibfnamefont {P.}~\bibnamefont {H{\"{a}}nggi}},\
  }\href@noop {} {\bibfield  {journal} {\bibinfo  {journal} {Phys. Rev. E}\
  }\textbf {\bibinfo {volume} {49}},\ \bibinfo {pages} {3626} (\bibinfo {year}
  {1994})}\BibitemShut {NoStop}%
\bibitem [{\citenamefont {MacLachlan}(1964)}]{maclachlan1964theory}%
  \BibitemOpen
  \bibfield  {author} {\bibinfo {author} {\bibfnamefont {N.~W.}\ \bibnamefont
  {MacLachlan}},\ }\href@noop {} {\emph {\bibinfo {title} {{Theory and
  Application of Mathieu Functions}}}},\ (\bibinfo  {publisher}
  {Dover Books, Dover},\ \bibinfo {year} {1964})\BibitemShut {NoStop}%
\bibitem [{\citenamefont {Nayfeh}\ and\ \citenamefont
  {Mook}(2008)}]{nayfeh2008nonlinear}%
  \BibitemOpen
  \bibfield  {author} {\bibinfo {author} {\bibfnamefont {A.~H.}\ \bibnamefont
  {Nayfeh}}\ and\ \bibinfo {author} {\bibfnamefont {D.~T.}\ \bibnamefont
  {Mook}},\ }\href@noop {} {\emph {\bibinfo {title} {{Nonlinear
  Oscillations}}}},\ Wiley Classics Library\ (\bibinfo  {publisher} {Wiley, New York},\
  \bibinfo {year} {2008})\BibitemShut {NoStop}%
\bibitem [{\citenamefont {Citro}\ \emph {et~al.}(2015)\citenamefont {Citro},
  \citenamefont {{Dalla Torre}}, \citenamefont {D'Alessio}, \citenamefont
  {Polkovnikov}, \citenamefont {Babadi}, \citenamefont {Oka},\ and\
  \citenamefont {Demler}}]{Citro2015}%
  \BibitemOpen
  \bibfield  {author} {\bibinfo {author} {\bibfnamefont {R.}~\bibnamefont
  {Citro}}, \bibinfo {author} {\bibfnamefont {E.~G.}\ \bibnamefont {{Dalla
  Torre}}}, \bibinfo {author} {\bibfnamefont {L.}~\bibnamefont {D'Alessio}},
  \bibinfo {author} {\bibfnamefont {A.}~\bibnamefont {Polkovnikov}}, \bibinfo
  {author} {\bibfnamefont {M.}~\bibnamefont {Babadi}}, \bibinfo {author}
  {\bibfnamefont {T.}~\bibnamefont {Oka}}, \ and\ \bibinfo {author}
  {\bibfnamefont {E.}~\bibnamefont {Demler}},\ }\href@noop {} {\bibfield
  {journal} {\bibinfo  {journal} {Ann. Phys. (N. Y).}\ }\textbf {\bibinfo
  {volume} {360}},\ \bibinfo {pages} {694} (\bibinfo {year}
  {2015})}\BibitemShut {NoStop}%
\bibitem [{\citenamefont {Zhu}\ \emph {et~al.}(2016)\citenamefont {Zhu},
  \citenamefont {Rexin},\ and\ \citenamefont {Mathey}}]{Zhu2016}%
  \BibitemOpen
  \bibfield  {author} {\bibinfo {author} {\bibfnamefont {B.}~\bibnamefont
  {Zhu}}, \bibinfo {author} {\bibfnamefont {T.}~\bibnamefont {Rexin}}, \ and\
  \bibinfo {author} {\bibfnamefont {L.}~\bibnamefont {Mathey}},\ }\href@noop {}
  {\bibfield  {journal} {\bibinfo  {journal} {Z. Naturforsch. A}\ }\textbf {\bibinfo {volume} {71}},\ \bibinfo {pages}
  {921} (\bibinfo {year} {2016})}\BibitemShut {NoStop}%
\bibitem [{\citenamefont {Kindt}\ and\ \citenamefont
  {Schmuttenmaer}(1999)}]{Kindt1999}%
  \BibitemOpen
  \bibfield  {author} {\bibinfo {author} {\bibfnamefont {J.~T.}\ \bibnamefont
  {Kindt}}\ and\ \bibinfo {author} {\bibfnamefont {C.~A.}\ \bibnamefont
  {Schmuttenmaer}},\ }\href@noop {} {\bibfield  {journal} {\bibinfo  {journal}
  {J. Chem. Phys.}\ }\textbf {\bibinfo {volume} {110}},\ \bibinfo {pages}
  {8589} (\bibinfo {year} {1999})}\BibitemShut {NoStop}%
\bibitem [{\citenamefont {N{\v{e}}mec}\ \emph {et~al.}(2002)\citenamefont
  {N{\v{e}}mec}, \citenamefont {Kadlec},\ and\ \citenamefont
  {Ku{\v{z}}el}}]{Nemec2002}%
  \BibitemOpen
  \bibfield  {author} {\bibinfo {author} {\bibfnamefont {H.}~\bibnamefont
  {N{\v{e}}mec}}, \bibinfo {author} {\bibfnamefont {F.}~\bibnamefont {Kadlec}},
  \ and\ \bibinfo {author} {\bibfnamefont {P.}~\bibnamefont {Ku{\v{z}}el}},\
  }\href@noop {} {\bibfield  {journal} {\bibinfo  {journal} {J. Chem. Phys.}\
  }\textbf {\bibinfo {volume} {117}},\ \bibinfo {pages} {8454} (\bibinfo {year}
  {2002})}\BibitemShut {NoStop}%
\bibitem [{\citenamefont {Orenstein}\ and\ \citenamefont
  {Dodge}(2015)}]{Orenstein2015}%
  \BibitemOpen
  \bibfield  {author} {\bibinfo {author} {\bibfnamefont {J.}~\bibnamefont
  {Orenstein}}\ and\ \bibinfo {author} {\bibfnamefont {J.~S.}\ \bibnamefont
  {Dodge}},\ }\href@noop {} {\bibfield  {journal} {\bibinfo  {journal} {Phys.
  Rev. B}\ }\textbf {\bibinfo {volume} {92}},\ \bibinfo {pages} {134507}
  (\bibinfo {year} {2015})}\BibitemShut {NoStop}%
\bibitem [{\citenamefont {Shao}\ \emph {et~al.}(2016)\citenamefont {Shao},
  \citenamefont {Tohyama}, \citenamefont {Luo},\ and\ \citenamefont
  {Lu}}]{Shao2016}%
  \BibitemOpen
  \bibfield  {author} {\bibinfo {author} {\bibfnamefont {C.}~\bibnamefont
  {Shao}}, \bibinfo {author} {\bibfnamefont {T.}~\bibnamefont {Tohyama}},
  \bibinfo {author} {\bibfnamefont {H.~G.}\ \bibnamefont {Luo}}, \ and\
  \bibinfo {author} {\bibfnamefont {H.}~\bibnamefont {Lu}},\ }\href@noop {}
  {\bibfield  {journal} {\bibinfo  {journal} {Phys. Rev. B}\ }\textbf {\bibinfo
  {volume} {93}},\ \bibinfo {pages} {195144} (\bibinfo {year}
  {2016})}\BibitemShut {NoStop}%
\bibitem [{\citenamefont {Kennes}\ \emph
  {et~al.}(2017{\natexlab{a}})\citenamefont {Kennes}, \citenamefont {Wilner},
  \citenamefont {Reichman},\ and\ \citenamefont {Millis}}]{Kennes2017b}%
  \BibitemOpen
  \bibfield  {author} {\bibinfo {author} {\bibfnamefont {D.~M.}\ \bibnamefont
  {Kennes}}, \bibinfo {author} {\bibfnamefont {E.~Y.}\ \bibnamefont {Wilner}},
  \bibinfo {author} {\bibfnamefont {D.~R.}\ \bibnamefont {Reichman}}, \ and\
  \bibinfo {author} {\bibfnamefont {A.~J.}\ \bibnamefont {Millis}},\
  }\href@noop {} {\bibfield  {journal} {\bibinfo  {journal} {Phys. Rev. B}\
  }\textbf {\bibinfo {volume} {96}},\ \bibinfo {pages} {054506} (\bibinfo
  {year} {2017}{\natexlab{a}})}\BibitemShut {NoStop}%
  \bibitem [{Note1()}]{Note1}%
  \BibitemOpen
  \bibinfo {note} {Importance of phase slips was discussed recently in Ref.
  \protect \rev@citealpnum {Kennes2017a}.}\BibitemShut {Stop}%
\bibitem [{\citenamefont {Coddington}\ and\ \citenamefont
  {Carlson}(1997)}]{coddington1997linear}%
  \BibitemOpen
  \bibfield  {author} {\bibinfo {author} {\bibfnamefont {A.}~\bibnamefont
  {Coddington}}\ and\ \bibinfo {author} {\bibfnamefont {R.}~\bibnamefont
  {Carlson}},\ }\href@noop {} {\emph {\bibinfo {title} {{Linear Ordinary
  Differential Equations}}}}\ (\bibinfo  {publisher} {Society for Industrial
  and Applied Mathematics, Philadelphia},\ \bibinfo {year} {1997})\BibitemShut {NoStop}%
\bibitem [{\citenamefont {Cesari}(2013)}]{cesari2013asymptotic}%
  \BibitemOpen
  \bibfield  {author} {\bibinfo {author} {\bibfnamefont {L.}~\bibnamefont
  {Cesari}},\ }\href@noop {} {\emph {\bibinfo {title} {{Asymptotic Behavior and
  Stability Problems in Ordinary Differential Equations}}}},\ Ergebnisse der
  Mathematik und ihrer Grenzgebiete. 2. Folge\ (\bibinfo  {publisher} {Springer, Berlin, Heidelberg},\ \bibinfo {year} {2013})\BibitemShut {NoStop}%
  \bibitem [{\citenamefont {Subedi}\ \emph {et~al.}(2014)\citenamefont {Subedi},
  \citenamefont {Cavalleri},\ and\ \citenamefont {Georges}}]{Subedi2014}%
  \BibitemOpen
  \bibfield  {author} {\bibinfo {author} {\bibfnamefont {A.}~\bibnamefont
  {Subedi}}, \bibinfo {author} {\bibfnamefont {A.}~\bibnamefont {Cavalleri}}, \
  and\ \bibinfo {author} {\bibfnamefont {A.}~\bibnamefont {Georges}},\
  }\href@noop {} {\bibfield  {journal} {\bibinfo  {journal} {Phys. Rev. B}\
  }\textbf {\bibinfo {volume} {89}},\ \bibinfo {pages} {220301} (\bibinfo
  {year} {2014})}\BibitemShut {NoStop}%
\bibitem [{Note2()}]{Note2}%
  \BibitemOpen
  \bibinfo {note} {In equilibrium, $\protect \operatorname {Im}\sigma (\omega =
  0.5 \omega _\protect \text {Jp1}) \approx 10 \ \Omega ^{-1} \protect \text
  {cm}^{-1}$.}\BibitemShut {Stop}%
\bibitem [{\citenamefont {Mitrano}\ \emph {et~al.}(2016)\citenamefont
  {Mitrano}, \citenamefont {Cantaluppi}, \citenamefont {Nicoletti},
  \citenamefont {Kaiser}, \citenamefont {Perucchi}, \citenamefont {Lupi},
  \citenamefont {{Di Pietro}}, \citenamefont {Pontiroli}, \citenamefont
  {Ricc{\`{o}}}, \citenamefont {Clark}, \citenamefont {Jaksch},\ and\
  \citenamefont {Cavalleri}}]{Mitrano2016}%
  \BibitemOpen
  \bibfield  {author} {\bibinfo {author} {\bibfnamefont {M.}~\bibnamefont
  {Mitrano}}, \bibinfo {author} {\bibfnamefont {A.}~\bibnamefont {Cantaluppi}},
  \bibinfo {author} {\bibfnamefont {D.}~\bibnamefont {Nicoletti}}, \bibinfo
  {author} {\bibfnamefont {S.}~\bibnamefont {Kaiser}}, \bibinfo {author}
  {\bibfnamefont {A.}~\bibnamefont {Perucchi}}, \bibinfo {author}
  {\bibfnamefont {S.}~\bibnamefont {Lupi}}, \bibinfo {author} {\bibfnamefont
  {P.}~\bibnamefont {{Di Pietro}}}, \bibinfo {author} {\bibfnamefont
  {D.}~\bibnamefont {Pontiroli}}, \bibinfo {author} {\bibfnamefont
  {M.}~\bibnamefont {Ricc{\`{o}}}}, \bibinfo {author} {\bibfnamefont {S.~R.}\
  \bibnamefont {Clark}}, \bibinfo {author} {\bibfnamefont {D.}~\bibnamefont
  {Jaksch}}, \ and\ \bibinfo {author} {\bibfnamefont {A.}~\bibnamefont
  {Cavalleri}},\ }\href@noop {} {\bibfield  {journal} {\bibinfo  {journal}
  {Nature (London)}\ }\textbf {\bibinfo {volume} {530}},\ \bibinfo {pages} {461}
  (\bibinfo {year} {2016})}\BibitemShut {NoStop}%
\bibitem [{\citenamefont {Knap}\ \emph {et~al.}(2016)\citenamefont {Knap},
  \citenamefont {Babadi}, \citenamefont {Refael}, \citenamefont {Martin},\ and\
  \citenamefont {Demler}}]{Knap2016}%
  \BibitemOpen
  \bibfield  {author} {\bibinfo {author} {\bibfnamefont {M.}~\bibnamefont
  {Knap}}, \bibinfo {author} {\bibfnamefont {M.}~\bibnamefont {Babadi}},
  \bibinfo {author} {\bibfnamefont {G.}~\bibnamefont {Refael}}, \bibinfo
  {author} {\bibfnamefont {I.}~\bibnamefont {Martin}}, \ and\ \bibinfo {author}
  {\bibfnamefont {E.}~\bibnamefont {Demler}},\ }\href@noop {} {\bibfield
  {journal} {\bibinfo  {journal} {Phys. Rev. B}\ }\textbf {\bibinfo {volume}
  {94}},\ \bibinfo {pages} {214504} (\bibinfo {year} {2016})}\BibitemShut
  {NoStop}%
\bibitem [{\citenamefont {Kim}\ \emph {et~al.}(2016)\citenamefont {Kim},
  \citenamefont {Nomura}, \citenamefont {Ferrero}, \citenamefont {Seth},
  \citenamefont {Parcollet},\ and\ \citenamefont {Georges}}]{Kim2016}%
  \BibitemOpen
  \bibfield  {author} {\bibinfo {author} {\bibfnamefont {M.}~\bibnamefont
  {Kim}}, \bibinfo {author} {\bibfnamefont {Y.}~\bibnamefont {Nomura}},
  \bibinfo {author} {\bibfnamefont {M.}~\bibnamefont {Ferrero}}, \bibinfo
  {author} {\bibfnamefont {P.}~\bibnamefont {Seth}}, \bibinfo {author}
  {\bibfnamefont {O.}~\bibnamefont {Parcollet}}, \ and\ \bibinfo {author}
  {\bibfnamefont {A.}~\bibnamefont {Georges}},\ }\href@noop {} {\bibfield
  {journal} {\bibinfo  {journal} {Phys. Rev. B}\ }\textbf {\bibinfo {volume}
  {94}},\ \bibinfo {pages} {155152} (\bibinfo {year} {2016})}\BibitemShut
  {NoStop}%
\bibitem [{\citenamefont {Komnik}\ and\ \citenamefont
  {Thorwart}(2016)}]{Komnik2016}%
  \BibitemOpen
  \bibfield  {author} {\bibinfo {author} {\bibfnamefont {A.}~\bibnamefont
  {Komnik}}\ and\ \bibinfo {author} {\bibfnamefont {M.}~\bibnamefont
  {Thorwart}},\ }\href@noop {} {\bibfield  {journal} {\bibinfo  {journal} {Eur.
  Phys. J. B}\ }\textbf {\bibinfo {volume} {89}},\ \bibinfo {pages} {244}
  (\bibinfo {year} {2016})}\BibitemShut {NoStop}%
\bibitem [{\citenamefont {Sentef}\ \emph {et~al.}(2016)\citenamefont {Sentef},
  \citenamefont {Kemper}, \citenamefont {Georges},\ and\ \citenamefont
  {Kollath}}]{Sentef2016}%
  \BibitemOpen
  \bibfield  {author} {\bibinfo {author} {\bibfnamefont {M.~A.}\ \bibnamefont
  {Sentef}}, \bibinfo {author} {\bibfnamefont {A.~F.}\ \bibnamefont {Kemper}},
  \bibinfo {author} {\bibfnamefont {A.}~\bibnamefont {Georges}}, \ and\
  \bibinfo {author} {\bibfnamefont {C.}~\bibnamefont {Kollath}},\ }\href@noop
  {} {\bibfield  {journal} {\bibinfo  {journal} {Phys. Rev. B}\ }\textbf
  {\bibinfo {volume} {93}},\ \bibinfo {pages} {144506} (\bibinfo {year}
  {2016})}\BibitemShut {NoStop}%
\bibitem [{\citenamefont {Kennes}\ \emph
  {et~al.}(2017{\natexlab{b}})\citenamefont {Kennes}, \citenamefont {Wilner},
  \citenamefont {Reichman},\ and\ \citenamefont {Millis}}]{Kennes2017}%
  \BibitemOpen
  \bibfield  {author} {\bibinfo {author} {\bibfnamefont {D.~M.}\ \bibnamefont
  {Kennes}}, \bibinfo {author} {\bibfnamefont {E.~Y.}\ \bibnamefont {Wilner}},
  \bibinfo {author} {\bibfnamefont {D.~R.}\ \bibnamefont {Reichman}}, \ and\
  \bibinfo {author} {\bibfnamefont {A.~J.}\ \bibnamefont {Millis}},\
  }\href@noop {} {\bibfield  {journal} {\bibinfo  {journal} {Nat. Phys.}\ }\textbf
  {\bibinfo {volume} {13}},\ \bibinfo {pages} {479}
  (\bibinfo {year} {2017}{\natexlab{b}})}\BibitemShut {NoStop}%
\bibitem [{\citenamefont {Murakami}\ \emph {et~al.}(2017)\citenamefont
  {Murakami}, \citenamefont {Tsuji}, \citenamefont {Eckstein},\ and\
  \citenamefont {Werner}}]{Murakami2017}%
  \BibitemOpen
  \bibfield  {author} {\bibinfo {author} {\bibfnamefont {Y.}~\bibnamefont
  {Murakami}}, \bibinfo {author} {\bibfnamefont {N.}~\bibnamefont {Tsuji}},
  \bibinfo {author} {\bibfnamefont {M.}~\bibnamefont {Eckstein}}, \ and\
  \bibinfo {author} {\bibfnamefont {P.}~\bibnamefont {Werner}},\ }\href@noop {}
  {\bibfield  {journal} {\bibinfo  {journal} {Phys. Rev. B}\ }\textbf {\bibinfo
  {volume} {96}},\ \bibinfo {pages} {045125} (\bibinfo {year}
  {2017})}\BibitemShut {NoStop}%
\bibitem [{\citenamefont {Mazza}\ and\ \citenamefont
  {Georges}(2017)}]{Mazza2017}%
  \BibitemOpen
  \bibfield  {author} {\bibinfo {author} {\bibfnamefont {G.}~\bibnamefont
  {Mazza}}\ and\ \bibinfo {author} {\bibfnamefont {A.}~\bibnamefont
  {Georges}},\ }\href@noop {} {\bibfield  {journal} {\bibinfo  {journal} {Phys.
  Rev. B}\ }\textbf {\bibinfo {volume} {96}},\ \bibinfo {pages} {064515}
  (\bibinfo {year} {2017})}\BibitemShut {NoStop}%
\bibitem [{\citenamefont {Babadi}\ \emph {et~al.}(2017)\citenamefont {Babadi},
  \citenamefont {Knap}, \citenamefont {Martin}, \citenamefont {Refael},\ and\
  \citenamefont {Demler}}]{Babadi2017}%
  \BibitemOpen
  \bibfield  {author} {\bibinfo {author} {\bibfnamefont {M.}~\bibnamefont
  {Babadi}}, \bibinfo {author} {\bibfnamefont {M.}~\bibnamefont {Knap}},
  \bibinfo {author} {\bibfnamefont {I.}~\bibnamefont {Martin}}, \bibinfo
  {author} {\bibfnamefont {G.}~\bibnamefont {Refael}}, \ and\ \bibinfo {author}
  {\bibfnamefont {E.}~\bibnamefont {Demler}},\ }\href@noop {} {\bibfield
  {journal} {\bibinfo  {journal} {Phys. Rev. B}\ }\textbf {\bibinfo {volume}
  {96}},\ \bibinfo {pages} {014512} (\bibinfo {year} {2017})}\BibitemShut
  {NoStop}%
\bibitem [{\citenamefont {Nava}\ \emph {et~al.}()\citenamefont {Nava},
  \citenamefont {Giannetti}, \citenamefont {Georges}, \citenamefont {Tosatti},\
  and\ \citenamefont {Fabrizio}}]{Nava2017}%
  \BibitemOpen
  \bibfield  {author} {\bibinfo {author} {\bibfnamefont {A.}~\bibnamefont
  {Nava}}, \bibinfo {author} {\bibfnamefont {C.}~\bibnamefont {Giannetti}},
  \bibinfo {author} {\bibfnamefont {A.}~\bibnamefont {Georges}}, \bibinfo
  {author} {\bibfnamefont {E.}~\bibnamefont {Tosatti}}, \ and\ \bibinfo
  {author} {\bibfnamefont {M.}~\bibnamefont {Fabrizio}},\ }\href@noop {}
  {\bibinfo  {journal} {arXiv:1704.05613}\ }\BibitemShut {NoStop}%
\bibitem [{\citenamefont {Kennes}\ and\ \citenamefont
  {Millis}(2017)}]{Kennes2017a}%
  \BibitemOpen
\bibfield  {journal} {  }\bibfield  {author} {\bibinfo {author} {\bibfnamefont
  {D.~M.}\ \bibnamefont {Kennes}}\ and\ \bibinfo {author} {\bibfnamefont
  {A.~J.}\ \bibnamefont {Millis}},\ }\href@noop {} {\bibfield  {journal}
  {\bibinfo  {journal} {Phys. Rev. B}\ }\textbf {\bibinfo {volume} {96}},\
  \bibinfo {pages} {064507} (\bibinfo {year} {2017})}\BibitemShut {NoStop}%
\end{thebibliography}
\end{document}